\documentclass[prx,aps,showpacs,twocolumn,superscriptaddress,longbibliography]{revtex4-1}
\usepackage{graphicx}
\usepackage{amsmath,amssymb,bbold,bm,color}
\usepackage{float}
\usepackage{epstopdf}
\usepackage{tabularx}
\usepackage{braket}
\usepackage{booktabs}
\usepackage{array}
\usepackage{blindtext}
\DeclareMathAlphabet\mathbfcal{OMS}{cmsy}{b}{n}
\usepackage[colorlinks=true,linkcolor=blue,anchorcolor=red,citecolor=blue,urlcolor=blue]{hyperref}
\allowdisplaybreaks[4]

\begin{document}

\title{A Clarification on Quantum-Metric-Induced Nonlinear Transport}

\author{Xiao-Bin Qiang}
\email{These authors contributed equally to this work.}
\affiliation{State Key Laboratory of Quantum Functional Materials, Department of Physics, and Guangdong Basic Research Center of Excellence for Quantum Science, Southern University of Science and Technology (SUSTech), Shenzhen 518055, China}
\affiliation{Quantum Science Center of Guangdong-Hong Kong-Macao Greater Bay Area (Guangdong), Shenzhen 518045, China}

\author{Tianyu Liu}
\email{These authors contributed equally to this work.}
\affiliation{Shenzhen Institute for Quantum Science and Engineering and Department of Physics, Southern University of Science and Technology (SUSTech), Shenzhen 518055, China}
\affiliation{Shenzhen Key Laboratory of Quantum Science and Engineering, Shenzhen 518055, China}

\author{Zi-Xuan Gao}
\email{These authors contributed equally to this work.}
\affiliation{State Key Laboratory of Quantum Functional Materials, Department of Physics, and Guangdong Basic Research Center of Excellence for Quantum Science, Southern University of Science and Technology (SUSTech), Shenzhen 518055, China}
\affiliation{Quantum Science Center of Guangdong-Hong Kong-Macao Greater Bay Area (Guangdong), Shenzhen 518045, China}

\author{Hai-Zhou Lu}
\email{Corresponding author: luhz@sustech.edu.cn}
\affiliation{State Key Laboratory of Quantum Functional Materials, Department of Physics, and Guangdong Basic Research Center of Excellence for Quantum Science, Southern University of Science and Technology (SUSTech), Shenzhen 518055, China}
\affiliation{Quantum Science Center of Guangdong-Hong Kong-Macao Greater Bay Area (Guangdong), Shenzhen 518045, China}

\author{X. C. Xie}
\affiliation{International Center for Quantum Materials, School of Physics, Peking University, Beijing 100871, China}
\affiliation{Interdisciplinary Center for Theoretical Physics and Information Sciences (ICTPIS), Fudan University, Shanghai 200433, China}
\affiliation{Hefei National Laboratory, Hefei 230088, China}

\begin{abstract}
Over the years, Berry curvature, which is associated with the imaginary part of the quantum geometric tensor, has profoundly impacted many branches of physics. Recently, quantum metric, the real part of the quantum geometric tensor, has been recognized as indispensable in comprehensively characterizing the intrinsic properties of condensed matter systems. The intrinsic second-order nonlinear conductivity induced by the quantum metric has attracted significant recent interest. However, its expression varies across the literature. Here, we reconcile this discrepancy by systematically examining the nonlinear conductivity using the standard perturbation theory, the wave packet dynamics, and the Luttinger-Kohn approach. Moreover, inspired by the Dirac model, we propose a toy model that suppresses the Berry-curvature-induced nonlinear transport, making it suitable for studying the quantum-metric-induced nonlinear conductivity. \textcolor{black}{This work provides a clearer and more unified understanding of the quantum-metric contributions to nonlinear transport. It also establishes a solid foundation for future theoretical developments and experimental explorations in this highly active and rapidly evolving field.}
\end{abstract}

\date{\today}
\maketitle

\section{Introduction} 
Berry curvature \cite{berry1984} characterizes the geometry of the Hilbert space and has significantly influenced the research paradigm of modern condensed matter physics \cite{xiaodi2010}. One prominent example is the family of quantum Hall effects. Both the integer quantum Hall effect \cite{klitzing1980} and the quantum anomalous Hall effect \cite{haldane1988} originate from the quantized integral of Berry curvature in the Brillouin zone \cite{thouless1982}. The uniform distribution of Berry curvature in the Landau levels responsible for the fractional quantum Hall effect \cite{wangjie2021} has made such uniformity a key design criterion for realizing the fractional quantum anomalous Hall effect \cite{parameswaran2012, roy2014, jackson2015}, which is a candidate platform for hosting Fibonacci anyons suitable for universal topological quantum computation \cite{liu2025}. In addition to various quantum Hall effects, Berry curvature also plays a key role in topological phases of matter \cite{hasan2010, armitage2018}, orbital magnetization \cite{xiaodi2005,xiaodi2006,thonhauser2005}, and nonlinear transport \cite{FuL15prl, Lu18prl, Lu19nc, Dimi20prl, Dimi22prl, ChenR24prb}.

Quantum metric is also a geometric measure and, together with Berry curvature, constitutes the quantum geometric tensor of Hilbert space \cite{Provost80cmp, AA90prl, Resta11epjb, Torma23prl, Lu24nsr}. The quantum metric is critical for understanding flat-band superconductivity \cite{Peotta15nc, Julku16prl, Torma17prb, Torma18prb, Torma22prb}, and is also essential for elucidating the fractional quantum anomalous Hall effect \cite{roy2014, jackson2015, wangjie2021, liuzhao2024} and nonlinear transport \cite{GaoY21prl, YangSY21prl, XuSY23science, GaoWB23nature}. Although analogous to Berry curvature in the context of nonlinear transport, it is crucial to note that quantum-metric-induced nonlinear transport is intrinsic (i.e., relaxation-time-independent) \cite{GaoY14prl, YanBH24prl, Dimi23prb}. Such nonlinear transport has recently been proposed and identified in antiferromagnets, where both the inversion symmetry ($\mathcal P$) and the time-reversal symmetry ($\mathcal T$) are broken but the combined $\mathcal{PT}$ is preserved (e.g., CuMnAs \cite{GaoY21prl}, Mn$_2$Au \cite{YangSY21prl}, and MnBi$_2$Te$_4$ \cite{XuSY23science,GaoWB23nature}), \textcolor{black}{further demonstrating a broad relevance and growing impact of this topic.}

Multiple theories have been proposed to elucidate the quantum-metric-induced nonlinear transport \cite{GaoY14prl,YanBH24prl,Dimi23prb}, with their predictions partly corroborated by transport experiments of MnBi$_2$Te$_4$ \cite{XuSY23science, GaoWB23nature}. However, there remains an inconsistency regarding the specific form of quantum-metric-induced nonlinear conductivity within these theories (see Table~\ref{Tab: Theory} for details). Moreover, the wave packet dynamics \cite{GaoY14prl} predicts the suppression of in-plane nonlinear transport under any out-of-plane $n$-fold rotational symmetry $\mathcal{C}_n^z$ ($n = 2, 3, 4, 6$) and the absence of longitudinal response \cite{GaoY21prl,YangSY21prl}, but the Luttinger-Kohn approach \cite{YanBH24prl} and the quantum kinetics \cite{Dimi23prb} are compatible with $\mathcal{C}_n^z$ and longitudinal response. This discrepancy aggravates the existing confusion in understanding quantum-metric-induced nonlinear transport. Therefore, a clarification of the form of quantum-metric-induced nonlinear conductivity is aspired.

In this paper, we aim to reconcile existing theoretical formulations of quantum-metric-induced nonlinear conductivity. Treating the driving electric field as a perturbation, we first derive the electric-field-modified Berry connection $\tilde{\mathbfcal{A}}_n(\mathbf k)$ and band energy $\tilde{\varepsilon}_{n\mathbf k}$ via the standard perturbation theory. A general expression for second-order nonlinear conductivity is derived. We then elucidate that the same nonlinear conductivity can be essentially derived through the wave packet dynamics, which yields identical electric-field-modified Berry connection and band energy. Moreover, the nonlinear conductivity obtained from the Luttinger-Kohn approach can be made consistent with those arising from the standard perturbation theory and the wave packet dynamics. Lastly, based on symmetry considerations, we propose a toy model that suppresses the Berry-curvature-induced second-order nonlinear transport, thereby highlighting the nonlinear transport resulting from the quantum metric.

\begin{table}[t]
\caption{Analytical expressions for the quantum-metric-induced second-order nonlinear conductivity $\sigma_{ijk}^{\text{qm}}$, where indices $i,j,k$ denote spatial  directions. The results are derived using the wave packet dynamics (WPD), the Luttinger-Kohn approach (LKA), and the quantum kinetics (QK). Here, $[d\mathbf k]\equiv d^d\mathbf k/(2\pi)^d$ with $d$ being the dimension, $f_0= f_0(\varepsilon_{n\mathbf k})$ represents the Fermi-Dirac distribution function evaluated at the unperturbed band energy $\varepsilon_{n\mathbf k}$ for band index $n$ and crystal momentum $\mathbf k$, and $\mathbfcal{G}_n= \mathbfcal{G}_n(\mathbf k)$ is the band-normalized quantum metric tensor of the $n$th band.} \label{Tab: Theory}
\centering
\setlength{\tabcolsep}{7pt} % weight
\renewcommand{\arraystretch}{2} % height
\begin{tabular}{cc}
\hline\hline
\textbf{Theories} & \textbf{Expressions} \\
\hline
WPD \cite{GaoY14prl} & $\displaystyle-\frac{e^3}{\hbar}\sum_n\int[d\mathbf k] \big[\partial_i\mathcal G_n^{jk}-\tfrac{1}{2}(\partial_k \mathcal G_n^{ij}+\partial_j \mathcal G_n^{ik})\big]f_0 $\\
LKA \cite{YanBH24prl} & $\displaystyle-\frac{e^3}{\hbar}\sum_n\int [d\mathbf k] \big [2\partial_i \mathcal G_n^{jk}-\tfrac{1}{2}(\partial_k \mathcal G_n^{ij}+\partial_j \mathcal G_n^{ik})\big ]f_0 $\\
QK \cite{Dimi23prb} & $\displaystyle-\frac{e^3}{\hbar}\sum_n\int [d\mathbf k]\big[\tfrac{1}{2}\partial_i \mathcal G_n^{jk}-(\partial_k \mathcal G_n^{ij}+\partial_j \mathcal G_n^{ik})\big]f_0 $ \\
\hline\hline
\end{tabular}
\end{table}

\section{Standard perturbation theory}
\label{sec2}
For a system subject to a weak driving electric field $\mathbf E$, the current density can be written as
\begin{equation} \label{Eq: J_def}
\mathbf{J}=-e\sum_n\int[d\mathbf{k}]\tilde{\mathbf{v}}_n(\mathbf k) f(\mathbf k),
\end{equation}
where $e$ is the elementary charge, $[d\mathbf{k} ]\equiv d^d\mathbf{k}/(2\pi)^d$ with $d$ denoting the dimension, $\tilde{\mathbf{v}}_n(\mathbf k)$ is the electric-field-modified velocity of the $n$th energy band at crystal momentum $\mathbf k$, and $f(\mathbf k)$ denotes the non-equilibrium distribution function. Specifically, the electric-field-modified velocity is given by $\tilde{\mathbf{v}}_n(\mathbf k)=\tfrac{1}{\hbar}\bm \nabla_{\mathbf k} \tilde{\varepsilon}_{n\mathbf k}+\tfrac{e}{\hbar} \mathbf E \times \tilde{\mathbf \Omega}_n(\mathbf k)$ \cite{xiaodi2010}, where $\tilde{\varepsilon}_{n\mathbf k}$ is the electric-field-modified energy of the $n$th band, and $\tilde{\mathbf \Omega}_n(\mathbf k)=\bm\nabla_{\mathbf k}\times \tilde{\mathbfcal A}_n(\mathbf k)$ represents the electric-field-modified Berry curvature of the $n$th energy band, arising from the electric-field-modified intraband Berry connection $\tilde{\mathbfcal A}_n(\mathbf k)$. On the other hand, under the relaxation time approximation, $f(\mathbf k)$ can be extracted through the Boltzmann equation as $f(\mathbf k)=\sum_{\nu=0}^\infty \left(\frac{\tau e}{\hbar}\mathbf {E}\cdot \bm \nabla_{\mathbf k} \right)^\nu f_0 ( {\tilde\varepsilon}_{n\mathbf k} )$ \cite{Haug08book,Lu19nc,Lu23prb}, where $\tau$ is the relaxation time, and $f_0( {\tilde\varepsilon}_{n\mathbf k})$ represents the Fermi-Dirac distribution function evaluated at ${\tilde\varepsilon}_{n\mathbf k}$. For a sufficiently weak $\mathbf E$, the $i$th component of the response current [Eq.~\eqref{Eq: J_def}] can be approximated by a power series in $\mathbf E$ as
\begin{equation} \label{J2}
J_i=\sigma_{ij}E_j + \sigma_{ijk} E_j E_k+\cdots,
\end{equation}
where $\sigma_{ij}$ is the linear conductivity and $\sigma_{ijk}$ is the second-order nonlinear conductivity (indices $i,j,k$ label spatial directions). According to Eq.~\eqref{Eq: J_def}, the determination of $\sigma_{ijk}$ requires expanding $\tilde{\mathbf{v}}_n(\mathbf k)$ and $f(\mathbf k)$ to the second order of $\mathbf E$. Therefore, it would be sufficient to estimate $\tilde{\mathbfcal A}_n(\mathbf k)$ and $\tilde{\varepsilon}_{n\mathbf k}$ to the first and second orders of $\mathbf E$, respectively.

To estimate $\tilde{\mathbfcal A}_n(\mathbf k)$, we consider the real-space Hamiltonian
\begin{equation} \label{Eq: Hami}
\hat{\mathcal H} = \hat{\mathcal H}_0 +e\mathbf E\cdot \hat{\mathbf r},
\end{equation}
where $\hat{\mathcal H}_0$ is the Hamiltonian characterizing a field-free periodic system and $e\mathbf E\cdot \hat{\mathbf r}$ is a perturbation induced by a weak applied electric field $\mathbf E$. The eigenvalue problem of $\hat{\mathcal H}_0$ can be solved by Bloch's theorem with the eigenenergy denoted as $\varepsilon_{n\mathbf k}$ and the eigenvector $|\psi_{n\mathbf k}(\mathbf r)\rangle = \mathrm e^{\mathrm i \mathbf k \cdot \mathbf r}|u_{n\mathbf k}(\mathbf r)\rangle$ comprising a plane wave $\mathrm e^{\mathrm i \mathbf k \cdot \mathbf r}$ and a unit-cell-periodic part $|u_{n\mathbf k}(\mathbf r)\rangle$. The eigenvectors of $\hat{\mathcal H}$ can then be perturbatively derived, to the first order of $\mathbf E$, as $|\tilde\psi_{n\mathbf k}(\mathbf r)\rangle=|\psi_{n\mathbf k}(\mathbf r)\rangle+\tfrac{V}{(2\pi)^d} \int d^d\mathbf k'\sum_{m\neq n} \langle \psi_{m\mathbf k'}(\mathbf r) |e\mathbf E\cdot \hat{\mathbf r}| \psi_{n\mathbf k}(\mathbf r) \rangle (\varepsilon_{n\mathbf k}-\varepsilon_{m\mathbf k'})^{-1} |\psi_{m\mathbf k'}(\mathbf r)\rangle$. The unit-cell-periodic part of $|\tilde\psi_{n\mathbf k}(\mathbf r)\rangle$ consequently becomes (detailed derivations in the Supplemental Material \cite{Supp})
\begin{equation} \label{eigvec}
|\tilde u_{n\mathbf k} (\mathbf r) \rangle= |u_{n\mathbf k}(\mathbf r)\rangle+ \sum_{m\neq n} \frac{ e\mathbf E \cdot \mathbfcal A_{mn}(\mathbf k) }{\varepsilon_ {n\mathbf k}-\varepsilon_{m\mathbf k}} |u_{m\mathbf k}(\mathbf r)\rangle,
\end{equation}
where $\mathbfcal A_{mn}(\mathbf k)=\langle u_{m\mathbf k}(\mathbf r)| \mathrm i\bm\nabla_{\mathbf k}|u_{n\mathbf k}(\mathbf r)\rangle$ is the field-free interband Berry connection. Accordingly, we can write the electric-field-modified Berry connection, to the first order of $\mathbf E$, as
\begin{equation}\label{Eq: An_tot}
\tilde{\mathbfcal A}_n(\mathbf k)=\frac{\langle \tilde u_{n\mathbf k}(\mathbf r)| \mathrm i\bm\nabla_{\mathbf k}|\tilde u_{n\mathbf k}(\mathbf r)\rangle}{\langle \tilde u_{n\mathbf k}(\mathbf r)| \tilde u_{n\mathbf k}(\mathbf r)\rangle} \simeq \mathbfcal A_n(\mathbf k)+\mathbf G_n(\mathbf k)\cdot \mathbf E,
\end{equation}
where $\mathbfcal A_n(\mathbf k) =\langle u_{n\mathbf k}(\mathbf r)| \mathrm i\bm\nabla_{\mathbf k}| u_{n\mathbf k}(\mathbf r)\rangle$ is the field-free Berry connection of the $n$th energy band \cite{xiaodi2010} and $\mathbf{G}_n(\mathbf k) = 2e \text{Re} \sum_{m\neq n} \mathbfcal{A}_{nm}(\mathbf k) \mathbfcal{A}_{mn}(\mathbf k) / (\varepsilon_{n\mathbf k} - \varepsilon_{m\mathbf k})$ is the Berry connection polarizability \cite{GaoY14prl}. It is worth noting that $\mathbf{G}_n(\mathbf k)$ is $U(1)$ gauge invariant and the electric-field-induced Berry connection $\mathbf G_n (\mathbf k) \cdot \mathbf E$ is observable \cite{Supp}. The electric-field-induced Berry connection originates from the quantum metric, because $\mathbf G_n (\mathbf k)$ differs from the band-normalized quantum metric $\mathbfcal{G}_n(\mathbf k) = 2 \text{Re} \sum_{m\neq n} \mathbfcal{A}_{nm}(\mathbf k) \mathbfcal{A}_{mn}(\mathbf k) / (\varepsilon_{n\mathbf k} - \varepsilon_{m\mathbf k})$ only by a multiplicative factor of $e$ \cite{YanBH24prl,Dimi23prb}, and the latter is related to the quantum metric $\mathbf{g}_n(\mathbf k) = \text{Re} \sum_{m\neq n} \mathbfcal{A}_{nm}(\mathbf k) \mathbfcal{A}_{mn}(\mathbf k)$ \cite{Provost80cmp,AA90prl,Resta11epjb,Torma23prl,Lu24nsr} through $\mathbfcal{G}_n(\mathbf k) = -\partial \mathbf{g}_n(\mathbf k) / \partial \varepsilon_{n\mathbf k}$. 
  
The evaluation of $\tilde \varepsilon_{n\mathbf k}$ in the context of the standard perturbation theory is a more subtle issue. Both the first-order energy correction $\langle\psi_{n\mathbf k}(\mathbf r)| e\mathbf E\cdot \hat {\mathbf r} | \psi_{n\mathbf k}(\mathbf r) \rangle$ and the second-order energy correction $\sum_{m\neq n} \langle\psi_{n\mathbf k}(\mathbf r)| e\mathbf E\cdot \hat {\mathbf r} |\psi_{m\mathbf k}(\mathbf r)\rangle [{e\mathbf E \cdot \mathbfcal A_{mn}(\mathbf k)}/(\varepsilon_ {n\mathbf k}-\varepsilon_{m\mathbf k})]$ formally diverge, because the matrix elements of $\hat {\mathbf r}$ are not well-defined in the basis of Bloch eigenvectors \cite{Supp}. To resolve the issue, we construct the following unit-cell-periodic Hamiltonian
\begin{equation} \label{ansatz}
\mathcal H(\mathbf k)= \mathcal H_0(\mathbf k)+e\mathbf E \cdot \mathrm i \bm\nabla_{\mathbf k},
\end{equation}
where $ \mathcal H_0(\mathbf k) = \mathrm e^{-\mathrm i \mathbf k \cdot \mathbf r} \hat{\mathcal H}_0 \mathrm e^{\mathrm i \mathbf k \cdot \mathbf r}$ is the Bloch Hamiltonian. It is straightforward to find that the $n$th eigenvector of $\mathcal H(\mathbf k)$, to the first order of $\mathbf E$, coincides with Eq.~\eqref{eigvec}. This validates Eq.~\eqref{ansatz} as an appropriate ansatz for describing a periodic system subject to an electric field. The perturbative energy corrections then become well-defined and the electric-field-modified band energy, to the second order of $\mathbf E$, reads \cite{Supp}
\begin{equation}\label{Eq: en_tot}
\begin{split}
\tilde{\varepsilon}_{n\mathbf k}= \varepsilon_{n\mathbf k} + \frac e 2 \mathbf E \cdot \mathbf G_n(\mathbf k) \cdot \mathbf E,
\end{split}
\end{equation}
where we have neglected the first-order energy correction $\langle u_{n\mathbf k}(\mathbf r)|e\mathbf E\cdot \mathrm i\bm\nabla_{\mathbf k} | u_{n\mathbf k}(\mathbf r) \rangle = e\mathbf E\cdot \mathbfcal A_n(\mathbf k)$ because it depends on gauge and hence does not contribute to physical observables. We mention that the first-order energy correction arises from the second term in Eq.~\eqref{ansatz}---a regularization that overcomes the ill-definedness of the position operator $\hat{\mathbf r}$ and restores the translational symmetry, but inevitably introduces gauge dependence as $\mathrm i \bm\nabla_{\mathbf k} \rightarrow\mathrm i \bm\nabla_{\mathbf k} - \bm \nabla_{\mathbf k}\phi_{\mathbf k}$, where $\phi_{\mathbf k}$ is the phase of the chosen $U(1)$ gauge.

Equations~\eqref{Eq: An_tot} and \eqref{Eq: en_tot} allow for expanding $\tilde{\mathbf{v}}_n(\mathbf k)$ and $f(\mathbf k)$ to the second order of $\mathbf E$. The $i$th component of the velocity, to the second order of $\mathbf E$, is given by \cite{Supp}
\begin{equation}\label{Eq: v_tot}
\begin{aligned}
\tilde{v}_n^i=&\frac{1}{\hbar}\partial_{i}\varepsilon_n+\frac{e}{2\hbar}(\Omega_n^{ij}E_j+\Omega_n^{ik}E_k)\\
&+\frac{e}{2\hbar}\big[3\partial_{i} G_n^{jk}-(\partial_j G_n^{ik} + \partial_k G_n^{ij})\big]E_jE_k,
\end{aligned}    
\end{equation}
where the argument $\mathbf k$ is omitted to prevent misguidance in subscripts/superscripts (e.g., $\varepsilon_{n\mathbf k}\rightarrow \varepsilon_{n}$ and $\partial_{k_i}\rightarrow \partial_{i}$) and Berry curvature tensor $\Omega_n^{ij}=\epsilon_{ijk}\Omega_n^k=\partial_i\mathcal A_n^j-\partial_j\mathcal A_n^i$ is introduced to symmetrize indices $\{j,k\}$. On the other hand, the non-equilibrium distribution function, to the second order of $\mathbf E$, reads \cite{Supp}
\begin{equation}\label{Eq: f}
f=f_0 + \frac{e}{2}G_n^{ij} E_iE_j f_0' + \frac{e\tau}{\hbar} E_i\partial_i  f_0 +  \frac{e^2\tau^2}{\hbar^2}E_iE_j\partial_i\partial_j  f_0,
\end{equation}
with $f= f(\mathbf k)$, $f_0= f_0(\varepsilon_{n\mathbf k})$, and $f_0'=\partial f_0(\varepsilon_{n\mathbf k})/\partial \varepsilon_{n\mathbf k}$. The second term in Eq.~\eqref{Eq: f} is the electric-field-induced correction to the Fermi-Dirac distribution function.

\begin{figure}[t]
\centering 
\includegraphics[width=0.48\textwidth]{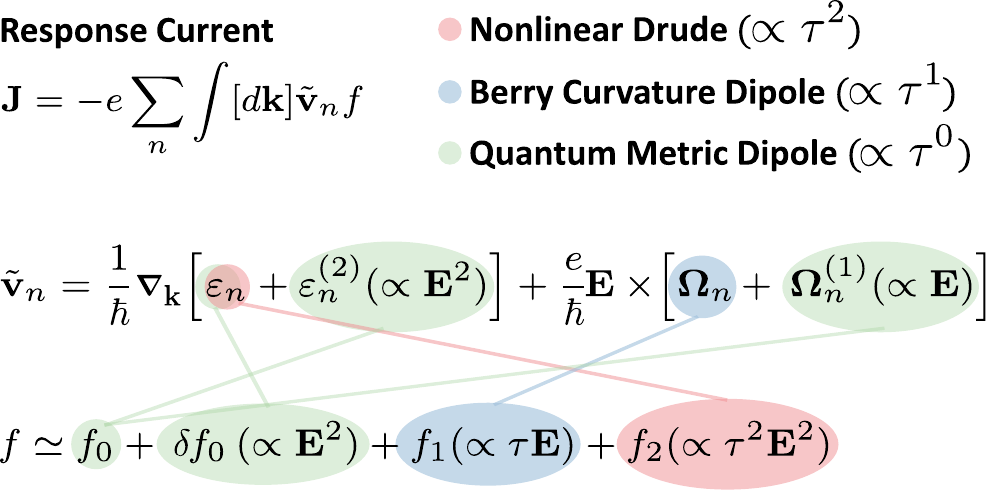}
\caption{Distinct physical origins of the second-order nonlinear conductivity. The contributions from the nonlinear Drude term, the Berry curvature dipole, and the quantum metric dipole are indicated in red, blue, and green, respectively. Note that $\mathbf E$ and $\mathbf E^2$ are used to demonstrate the order of electric-field-induced perturbation, and the exact $\mathbf E$ dependences of the relevant quantities are given in Eqs.~\eqref{Eq: An_tot}, \eqref{Eq: en_tot}, and \eqref{Eq: f}.
}
\label{Fig: demo}
\end{figure}

Plugging Eqs.~\eqref{Eq: v_tot} and~\eqref{Eq: f} into Eq.~\eqref{Eq: J_def}, the second-order nonlinear conductivity is obtained as \cite{Supp}
\begin{equation} \label{sigma}
\sigma_{ijk}=\sigma_{ijk}^{\text d}+\sigma_{ijk}^{\text{bc}}+\sigma_{ijk}^{\text{qm}},
\end{equation}
which originates from three distinct physical mechanisms, differentiated by their dependence on the relaxation time $\tau$: the Drude term $\sigma_{ijk}^{\text d}$ exhibits a quadratic $\tau$ dependence, the Berry-curvature-induced term $\sigma_{ijk}^{\text{bc}}$ scales linearly with $\tau$, and the quantum-metric-induced term $\sigma_{ijk}^{\text{qm}}$ is independent of $\tau$, thus being an intrinsic property (see Fig. \ref{Fig: demo}). The three contributions explicitly read 
\begin{align}
&\sigma_{ijk}^{\text{d}}=-\frac{e^3\tau^2}{\hbar^3}\sum_n\int [d\mathbf{k}]  \big( \partial_i \partial_j  \partial_k\varepsilon_n\big)f_0,
\\
&\sigma_{ijk}^{\text{bc}}=\frac{e^3\tau}{2\hbar^2}\sum_n\int [d\mathbf{k}]\big(\partial_k\Omega_n^{ij}+\partial_j\Omega_n^{ik}\big)f_0,
\\
&\sigma_{ijk}^{\text{qm}}=-\frac{e^3}{\hbar}\sum_n\int [d\mathbf{k}]\left[\partial_i\mathcal{G}_n^{jk}-\frac{1}{2}\left(\partial_j \mathcal{G}_n^{ik}+\partial_k \mathcal{G}_n^{ij}\right)\right]f_0,  \label{Eq: sigma}
\end{align}
where the band-normalized quantum metric is used in Eq.~\eqref{Eq: sigma} to facilitate comparison with existing theories (see Table.~\ref{Tab: Theory}). The quantum-metric-induced nonlinear conductivity [Eq.~\eqref{Eq: sigma}] constitutes the central result of the standard perturbation theory, consistent with that derived from the wave packet dynamics \cite{GaoY14prl} while distinct from those derived by the Luttinger-Kohn approach \cite{YanBH24prl} and the quantum kinetics \cite{Dimi23prb}. Alternatively, it can be reformulated in terms of $f_0'$ as
\begin{equation}\label{Eq: sigma_FS}
\sigma_{ijk}^{\text{qm}}=e^3\sum_n\int [d\mathbf k] \Lambda_n^{ijk} f_0',
\end{equation}
where $\Lambda_n^{ijk} = v_n^i\mathcal G_n^{jk}-\tfrac{1}{2}(v_n^j \mathcal G_n^{ik}+v_n^k \mathcal G_n^{ij})$ is known as the quantum metric dipole. According to Eq.~\eqref{Eq: sigma_FS}, it is straightforward to check that the quantum-metric-induced nonlinear longitudinal transport is prohibited, as can be seen from the identity $\sigma_{iii}^{\text{qm}}=0$. 

\section{Wave packet dynamics}
\label{sec3}
We now examine the consistency between the standard perturbation theory and the wave packet dynamics. Within the framework of the wave packet dynamics, the quantum state associated with the $n$th band of the Bloch Hamiltonian $\mathcal H_0(\mathbf k)$ is described by a wave packet $|W_n(\mathbf r) \rangle = \int  d^d\mathbf k w_n (\mathbf k) |\psi_{n\mathbf k}(\mathbf r)\rangle$, where the superposition weight $w_n (\mathbf k)$ satisfies $|w_n (\mathbf k)|^2=\tfrac{V}{(2\pi)^d}\delta(\mathbf k-\mathbf k_c)$ with $\mathbf k_c$ labeling the momentum center of the wave packet \cite{xiaodi2010}. In the presence of an applied electric field, the wave packet develops interband mixing and is thus modified to \cite{GaoY14prl}
\begin{equation}
\begin{aligned}
\left| \tilde{W}_n(\mathbf r) \right\rangle= &\int  d^d\mathbf k (1-\delta) w_n (\mathbf k) |\psi_{n\mathbf k}(\mathbf r)\rangle\\
& +\int  d^d\mathbf k \sum_{m\neq n} w_m^{(1)}(\mathbf k)|\psi_{m\mathbf k}(\mathbf r)\rangle,
\end{aligned}
\end{equation}
where the coefficient $w_m^{(1)}(\mathbf k)=w_n(\mathbf k) e\mathbf E\cdot \mathbfcal A_{mn}(\mathbf k)/(\varepsilon_{n\mathbf k}-\varepsilon_{m\mathbf k})$ represents the first-order amplitude of the admixture from the $m$th band and the parameter $\delta=\tfrac{1}{2}\tfrac{(2\pi)^d}{V} \int d^d\mathbf k \sum_{m\neq n}|w_m^{(1)}(\mathbf k)|^2$ ensures the normalization of the modified wave packet \cite{Supp}.

The electric-field-induced Berry connection can be extracted from the positional shift of the wave packet. Specifically, in the presence of the applied electric field, the wave packet center is located at \cite{Supp}
\begin{equation} \label{pos_shift}
\left\langle \tilde W_n(\mathbf r) \right| \hat{\mathbf r} \left| \tilde W_n(\mathbf r) \right\rangle= \mathbf{r}_c + \mathbf G_n(\mathbf k_c) \cdot \mathbf E,
\end{equation}
where $\mathbf{r}_c=\langle W_n(\mathbf r) | \hat{\mathbf r} | W_n(\mathbf r) \rangle=-\bm\nabla_{\mathbf{k}}\arg w_n(\mathbf k)|_{\mathbf{k=\mathbf{k}_c}}+\mathbfcal{A}_n(\mathbf k_c)$ represents the field-free wave packet center. The second term in Eq.~\eqref{pos_shift} can thus be understood as a correction to the Berry connection \cite{GaoY14prl}. Consequently, the electric-field-modified Berry connection reads $\tilde {\mathbfcal{A}}_n(\mathbf k_c)=\mathbfcal{A}_n(\mathbf k_c)+\mathbf G_n(\mathbf k_c) \cdot \mathbf E$, which is consistent with Eq.~\eqref{Eq: An_tot}. On the other hand, the electric-field-modified band energy can be calculated by extracting the gauge invariant part of $\langle \tilde W_n(\mathbf r)| \hat{\mathcal{H}} |\tilde W_n(\mathbf r) \rangle$ as $\tilde {\varepsilon}_{n\mathbf k_c}=\varepsilon_{n\mathbf k_c}+\tfrac{e}{2}\mathbf E \cdot \mathbf G_n(\mathbf k_c) \cdot \mathbf E$ \cite{Supp}, which is consistent with Eq.~\eqref{Eq: en_tot}. As the second-order nonlinear conductivity is fully determined by the electric-field-modified Berry connection (to the first order of $\mathbf E$) and the electric-field-modified band energy (to the second order of $\mathbf E$), the wave packet dynamics yields a nonlinear conductivity identical to that derived from the standard perturbation theory.

\section{Luttinger-Kohn approach}
\label{sec4}
Beyond the standard perturbation theory and the wave packet dynamics, the electric-field-modified Berry connection and band energy can also be extracted using the Luttinger-Kohn approach \cite{Luttinger55pr, Schrieffer66pr}. To implement the approach, we begin by rewriting Eq.~\eqref{ansatz} as 
\begin{equation} 
\mathcal H(\mathbf k)=\mathcal H_0(\mathbf k)+\lambda e\mathbf E\cdot\mathrm i\bm\nabla_{\mathbf k},
\end{equation}
where $\lambda$ is introduced for transparency to track the order of perturbation and can be set to unity as needed. We then perform a unitary Schrieffer-Wolff transformation $|u_{n\mathbf k}(\mathbf r)\rangle \rightarrow \mathrm e^{\lambda\mathcal S(\mathbf k)} |u_{n\mathbf k}(\mathbf r)\rangle$, where the generator $\mathcal S(\mathbf k)$ must be anti-Hermitian to guarantee the unitarity of the transformation \cite{Schrieffer66pr}. One of the key findings of the Luttinger-Kohn approach is that an appropriate choice of the generator $\mathcal S(\mathbf k)$ renders $\mathrm e^{\lambda\mathcal S(\mathbf k)} |u_{n\mathbf k}(\mathbf r)\rangle$ an eigenvector of $\mathcal H(\mathbf k)$ to the desired order \cite{Luttinger55pr, Schrieffer66pr} and the corresponding eigenenergies can be obtained from the eigenvalue problem of the transformed Hamiltonian $\mathcal H_\text{eff}(\mathbf k) = \mathrm e^{-\lambda \mathcal S(\mathbf k)} \mathcal H(\mathbf k) \mathrm e^{\lambda \mathcal S(\mathbf k)}$. To the second order of $\lambda$, the transformed Hamiltonian reads
\begin{equation} \label{Heff}
\begin{split}
\mathcal H_\text{eff}(\mathbf k)=\mathcal{H}_0(\mathbf k)+\lambda\big(e\mathbf E\cdot\mathrm i\bm\nabla_{\mathbf k}+ [\mathcal{H}_0(\mathbf k), \mathcal{S}(\mathbf k)]\big)
\\
+ \lambda^2\left([e\mathbf E\cdot\mathrm i\bm\nabla_{\mathbf k}, \mathcal{S}(\mathbf k)]+\frac{1}{2}[[\mathcal{H}_0(\mathbf k), \mathcal{S}(\mathbf k)], \mathcal{S}(\mathbf k)]  \right).
\end{split}
\end{equation}
The goal here is to choose a generator that eliminates the off-diagonal elements of the second term in Eq.~\eqref{Heff} such that $\mathcal H_\text{eff}(\mathbf k)$ becomes diagonalized to the first order of $\lambda$ and the energy correction arising from the third term (scaled as $\lambda^2$) coincides with its expectation value in $|u_{n\mathbf k}(\mathbf r)\rangle$. The elimination condition, which is referred to as the Luttinger-Kohn condition, gives rise to the off-diagonal elements of the generator as \cite{Supp}
\begin{equation} \label{S}
\mathcal S_{mn}(\mathbf k) = \frac{e\mathbf E\cdot \mathbfcal A_{mn}(\mathbf k)}{\varepsilon_{n\mathbf k}-\varepsilon_{m\mathbf k}}, 
\end{equation}
where $m\neq n$. It is worth noting that the diagonal entries of $\mathcal S(\mathbf k)$ cannot be uniquely determined, though the anti-Hermicity requires their real parts to vanish. Nevertheless, the $n$th eigenenergy of $\mathcal H(\mathbf k)$ coincides with $ \langle u_{n\mathbf k}(\mathbf r)| \mathcal H_\text{eff}(\mathbf k)| u_{n\mathbf k}(\mathbf r) \rangle$. Its gauge invariant part, at $\lambda=1$, corresponds to the electric-field-modified band energy and reads $\tilde{\varepsilon}_{n\mathbf k} = \varepsilon_{n\mathbf k} + \tfrac{e}{2} \mathbf E \cdot \mathbf G_n(\mathbf k) \cdot \mathbf E$ \cite{Supp}, which is identical to Eq.~\eqref{Eq: en_tot}. With the Schrieffer-Wolff transformed unit-cell-periodic state $\mathrm e^{\mathcal S(\mathbf k)} |u_{n\mathbf k}(\mathbf r)\rangle$, the electric-field-modified Berry connection reads $\tilde{\mathbfcal{A}}_n(\mathbf k)= \langle u_{n\mathbf k}(\mathbf r)| \mathrm e^{-\mathcal S(\mathbf k)} \mathrm i \bm\nabla_{\mathbf k} \mathrm e^{\mathcal S(\mathbf k)}| u_{n\mathbf k}(\mathbf r) \rangle =\mathbfcal A_n(\mathbf k)+\mathbf G_n(\mathbf k) \cdot \mathbf E$ \cite{Supp}, which is identical to Eq.~\eqref{Eq: An_tot}. Since the Luttinger-Kohn approach produces the same electric-field-modified Berry connection and band energy as the standard perturbation theory, it is expected to yield an identical second-order nonlinear conductivity.

\section{Origin of inconsistency}
\label{sec5}
Our derivations using three distinct methods (i.e., the standard perturbation theory in Sec.~\ref{sec2}, the wave packet dynamics in Sec.~\ref{sec3}, and the Luttinger-Kohn approach in Sec.~\ref{sec4}) agree with Ref.~\cite{GaoY14prl} but are inconsistent with Refs.~\cite{YanBH24prl} and~\cite{Dimi23prb}. Herein, we briefly discuss the origin of the inconsistency concerning the form of the quantum-metric-induced nonlinear conductivity.

\subsection{Luttinger-Kohn approach}
We note that our derivations in Sec.~\ref{sec4} differ from those of Ref.~\cite{YanBH24prl} in both the electric-field-modified band energy and the argument of the distribution function.

In the Luttinger-Kohn approach of Ref.~\cite{YanBH24prl}, the variation of band energy due to the applied electric field $\mathbf E$ (cf. Eq.~(5) of Ref.~\cite{YanBH24prl}) is attributed solely to the electric-field-induced Berry connection $\mathbf G_n(\mathbf k)\cdot \mathbf E$. However, it is crucial to recognize that the bare band energy itself must also be renormalized by the electric field, because the wave function [Eq.~\eqref{eigvec}] now acquires $\mathbf E$ dependence. Taking into account the $\mathbf E$ dependence, the bare band energy becomes ${\langle\tilde u_{n\mathbf k}(\mathbf r)|\mathcal H_0(\mathbf k) |\tilde u_{n\mathbf k}(\mathbf r)\rangle}/{\langle\tilde u_{n\mathbf k}(\mathbf r)|\tilde u_{n\mathbf k}(\mathbf r)\rangle} = \varepsilon_{n\mathbf k}-\tfrac{e}{2}\mathbf E \cdot \mathbf G_n(\mathbf k)\cdot \mathbf E$, which yields the correct electric-field-modified band energy [Eq.~\eqref{Eq: en_tot}] when combined with the Berry connection contribution $\mathbf E\cdot \mathbf G_n(\mathbf k)\cdot \mathbf E$.

When solving for the nonlinear conductivity within the Boltzmann formalism, Ref.~\cite{YanBH24prl} evaluates the Fermi-Dirac distribution function at the unrenormalized bare band energy $\varepsilon_{n\mathbf k}$. While this is justified in the context of linear response theory as the first-order energy correction $e\mathbf E\cdot\mathbfcal A_n(\mathbf k)$ depends on gauge and should not affect the observable physics, it is essential to expand the Fermi-Dirac distribution function to the second order of $\mathbf E$ when deriving the quantum-metric-induced nonlinear conductivity. Consequently, it is more appropriate to express the Fermi-Dirac distribution function as $f_0(\tilde \varepsilon_{n\mathbf k})=f_0(\varepsilon_{n\mathbf k})+\frac{e}{2} \mathbf E \cdot \mathbf G_n(\mathbf k) \cdot \mathbf E \tfrac{\partial f_0}{\partial \varepsilon_{n\mathbf k}}$, which incorporates an additional Fermi-sea occupation correction quadratic in $\mathbf E$ \cite{YangSY22prb, WangJ24prb, YangSY24arXiv}.

\subsection{Quantum kinetics}
Regarding the quantum-metric-induced nonlinear conductivity derived from quantum kinetics \cite{Dimi23prb}, the discrepancy in the expression (see Table.~\ref{Tab: Theory}) likely has a more complex and subtle origin. Although we do not attempt to reproduce the full density matrix formalism of Ref.~\cite{Dimi23prb}, we note that a so-called \emph{modified scattering time} $\tau/2$, where $\tau$ is the regular scattering time, is adopted when evaluating certain density matrix contributions. This modification of the relaxation time seems not adequately justified and may compromise the reliability of the derived quantum-metric-induced nonlinear conductivity (Table.~\ref{Tab: Theory}), which indeed differs from ours [Eq.~\eqref{Eq: sigma}].

\section{Toy model for quantum-metric-induced nonlinear transport}
We now construct a toy---though not necessarily minimal---model that is suitable for investigating the quantum-metric-induced second-order nonlinear transport. For simplicity, we restrict the analysis to a planar geometry such that only nonlinear Hall conductivities $\sigma_{xyy}^{\text{qm}}$ and $\sigma_{yxx}^{\text{qm}}$ are relevant. Symmetry analysis indicates that the emergence of nonlinear Hall conductivities requires the breaking of $\mathcal{C}_n^z$, $\mathcal P$, and $\mathcal T$ symmetries \cite{YangSY21prl}. In addition, $\mathcal P\mathcal T$ symmetry is preferred because it suppresses the nonlinear Hall effect arising from the Berry curvature dipole, rendering the quantum-metric-dipole contribution dominant. With all these requirements, we propose the following Bloch Hamiltonian
\begin{equation}\label{Eq: model}
\mathcal{H}_0(\mathbf k) = v \tau_x(k_x\sigma_x + k_y\sigma_y) + (m-bk^2) \tau_z+tk_x\tau_z,
\end{equation}
where $v$, $m$, $b$ and $t$ are model parameters; and $\bm\sigma$ and $\bm \tau$ are Pauli matrices. The first two terms in Eq.~\eqref{Eq: model} constitute a Dirac model with $\mathcal{PT}$ symmetry \cite{Shen17book}, whereas the last term, while preserving the $\mathcal{PT}$ symmetry, breaks $\mathcal{C}_n^z$, $\mathcal{P}$, and $\mathcal{T}$ symmetries.

\begin{figure}[t]
\centering 
\includegraphics[width=0.48\textwidth]{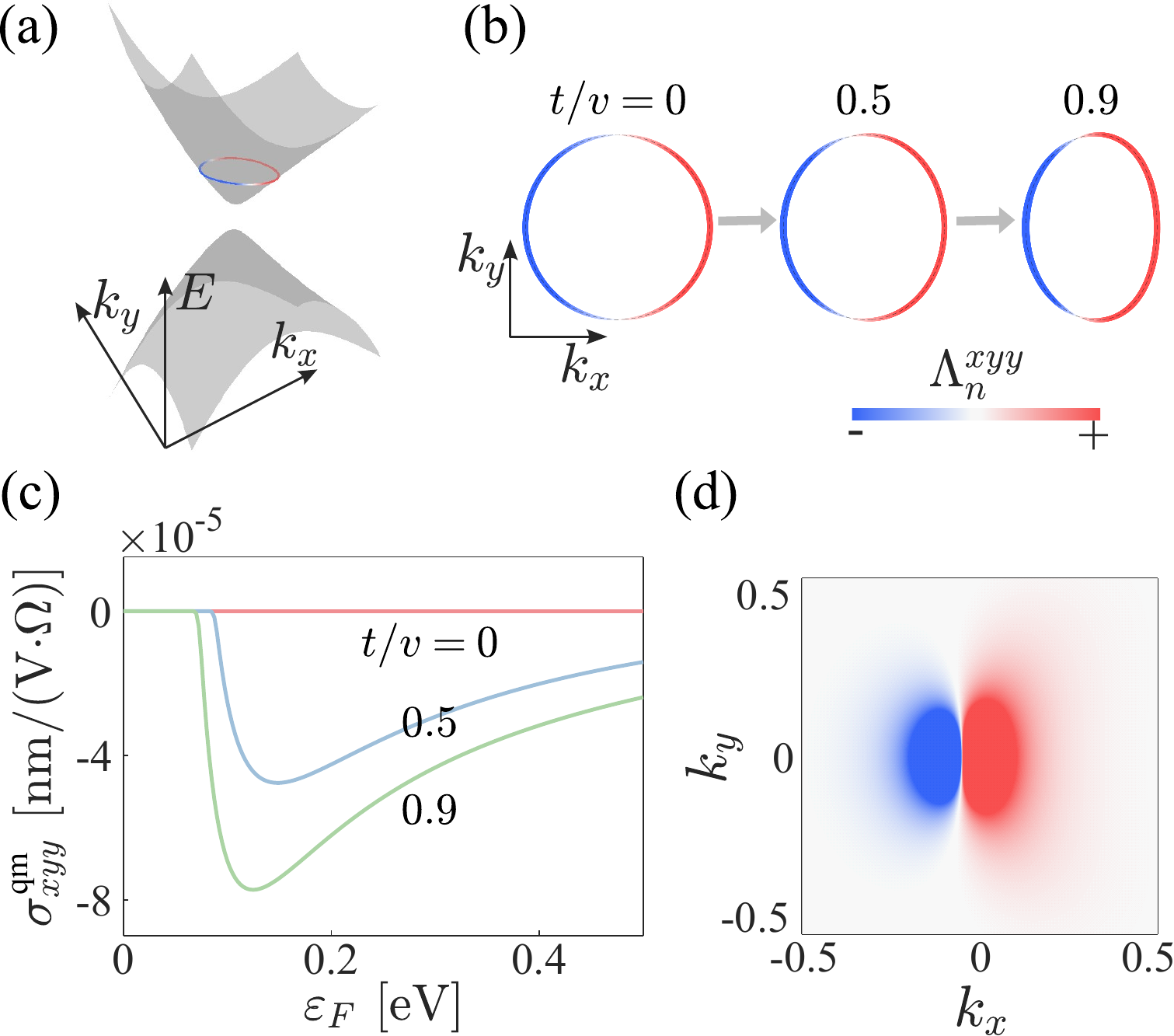}
\caption{Theoretical results from the toy model [Eq.~\eqref{Eq: model}]. (a) Band structure at $t/v=0.9$ with the Fermi surface placed at $\varepsilon_F=0.25$ eV. (b) Fermi surfaces at $\varepsilon_F=0.25$ eV for $t/v=0, 0.5, 0.9$. The color scale indicates the magnitude of the quantum metric dipole $\Lambda_n^{xyy}$. (c) Calculated nonlinear Hall conductivity $\sigma_{yxx}^{\text{qm}}$ as a function of the Fermi energy $\varepsilon_F$. (d) Momentum-resolved distribution of $\Lambda_n^{xyy}$ for $t/v=0.9$. The model parameters used are $v=1$ eV$\cdot$nm, $m=0.1$ eV, and $b=1$ eV$\cdot$nm$^2$.
}
\label{Fig: model}
\end{figure}

The Hamiltonian exhibits a deformed gapped Dirac cone band structure $\varepsilon_{\mathbf k}=\pm\sqrt{v^2k^2+(m-bk^2+tk_x)^2}$ [Fig.~\ref{Fig: model}(a)], where the parameter $t$ measures the degree of deformation. The evolution of the Fermi surface with respect to $t$ is illustrated in Fig. \ref{Fig: model}(b). For $t=0$, the symmetry-breaking term vanishes, resulting in a symmetric distribution of the quantum metric dipole $\Lambda_n^{xyy}$ on the Fermi surface [Fig.~\ref{Fig: model}(b)] and a vanishing quantum-metric-induced nonlinear Hall conductivity $\sigma_{xyy}^{\text{qm}}$ [red line, Fig.~\ref{Fig: model}(c)]. As $t$ increases, the asymmetry in the distribution of $\Lambda_n^{xyy}$ grows [Fig.~\ref{Fig: model}(b)], leading to enhanced nonlinear Hall conductivities [blue and green curves, Fig.~\ref{Fig: model}(c)]. To better visualize the asymmetry of $\Lambda_n^{xyy}$ across the Fermi surface, $\mathbf k$-resolved $\Lambda_n^{xyy}$ is shown in Fig.~\ref{Fig: model}(d).

\section{Conclusions} We have clarified the discrepancies among several existing theories for quantum-metric-induced second-order nonlinear transport by examining the nonlinear conductivity using three distinct approaches: the standard perturbation theory, the wave packet dynamics, and the Luttinger-Kohn approach. Through careful symmetry design, we have proposed a $\mathcal P \mathcal T$ symmetric toy model that suppresses the Berry-curvature-induced nonlinear Hall effect while allowing the quantum-metric-induced contribution to dominate. The magnitude of the resulting quantum-metric-induced nonlinear Hall conductivity positively correlates with the degree of breaking of rotational, inversion, and time-reversal symmetries. This work not only clarifies the inconsistency in the theory of nonlinear transport but also proposes a suitable model for studying the quantum-metric-induced nonlinear Hall effect, which can potentially be realized in quantum materials. Furthermore, our framework can be extended to incorporate disorder effects and higher-order quantum geometric contributions, paving the way for a more comprehensive understanding of nonlinear transport phenomena.
\\

\noindent \textit{Note added.---}When preparing the revised manuscript, we became aware of a related preprint \cite{Schnyder25arXiv}, in which the interband quantum-metric-induced nonlinear conductivity is consistent with our Eq.~\eqref{Eq: sigma}.

\begin{acknowledgments}
The authors are indebted to Q. Niu, Y. Gao, X. Yang, H. Lin, and Y. Chen for the insightful discussions. This work is supported by the National Key R\&D Program of China (2022YFA1403700), Innovation Program for Quantum Science and Technology (2021ZD0302400), the National Natural Science Foundation of China (12304196, 12350402, 12574173, and 12525401), Guangdong Basic and Applied Basic Research Foundation (2022A1515111034 and 2023B0303000011), Guangdong Provincial Quantum Science Strategic Initiative (GDZX2201001 and GDZX2401001), the Science, Technology and Innovation Commission of Shenzhen Municipality (ZDSYS20190902092905285), High-level Special Funds (G03050K004), the New Cornerstone Science Foundation through the XPLORER PRIZE, and the Center for Computational Science and Engineering of SUSTech.
\end{acknowledgments}

\bibliographystyle{apsrev4-1.bst}
\bibliography{QG}

\begin{thebibliography}{13}
\bibitem[{not({\natexlab{a}})}]{note1}
\bibinfo{note}{Consider a gauge transformation $|u_{n\mathbf k}(\mathbf r)
  \rangle \rightarrow e^{\mathrm i\phi_{n\mathbf k}}|u_{n\mathbf k}(\mathbf r)
  \rangle$. The resulting transformed interband Berry connection becomes
  $\mathbfcal A_{mn}(\mathbf k) \rightarrow \langle u_{m\mathbf k}(\mathbf r)|
  e^{-\mathrm i\phi_{m\mathbf k}} (\mathrm i\bm\nabla_{\mathbf k} e^{\mathrm
  i\phi_{n\mathbf k}}) |u_{n\mathbf k}(\mathbf r) \rangle + \langle u_{m\mathbf
  k}(\mathbf r)| e^{-\mathrm i\phi_{m\mathbf k}} e^{\mathrm i\phi_{n\mathbf k}}
  \mathrm i\bm\nabla_{\mathbf k} |u_{n\mathbf k}(\mathbf r) \rangle =
  e^{\mathrm i(\phi_{n\mathbf k}-\phi_{m\mathbf k})} \mathbfcal A_{mn}(\mathbf
  k)$, where we have used $\langle u_{m\mathbf k}(\mathbf r)| u_{n\mathbf
  k}(\mathbf r)\rangle = 0$ for $m\neq n$. Consequently, $\mathbfcal
  A_{nm}(\mathbf k)\mathbfcal A_{mn}(\mathbf k)$ possesses gauge invariance,
  which is inherited by the Berry connection polarizability.}
\bibitem[{not({\natexlab{b}})}]{note2}
\bibinfo{note}{To see how the band-normalized quantum metric $\mathbfcal
  G_n(\mathbf k)$ is related to the quantum metric $\mathbf g_n(\mathbf k)$, it
  would be helpful to first rewrite $\mathbfcal G_n(\mathbf k)$ and $\mathbf
  g_n(\mathbf k)$. We note that $\langle u_{m\mathbf k}(\mathbf r)| \mathcal
  H_0(\mathbf k) |u_{n\mathbf k}(\mathbf r)\rangle=0$ for $m \neq n$. This
  implies $\langle \bm \nabla_{\mathbf k} u_{m\mathbf k}(\mathbf r)| \mathcal
  H_0(\mathbf k) |u_{n\mathbf k}(\mathbf r)\rangle + \langle u_{m\mathbf
  k}(\mathbf r)| \mathcal H_0(\mathbf k) | \bm \nabla_{\mathbf k} u_{n\mathbf
  k}(\mathbf r)\rangle + \langle u_{m\mathbf k}(\mathbf r)| \bm \nabla_{\mathbf
  k} \mathcal H_0(\mathbf k) |u_{n\mathbf k}(\mathbf r)\rangle
  =(\varepsilon_{m\mathbf k}-\varepsilon_{n\mathbf k})\langle u_{m\mathbf
  k}(\mathbf r)| \bm \nabla_{\mathbf k} |u_{n\mathbf k}(\mathbf r)\rangle +
  \langle u_{m\mathbf k}(\mathbf r)| \bm \nabla_{\mathbf k} \mathcal
  H_0(\mathbf k) |u_{n\mathbf k}(\mathbf r)\rangle =0$, from which we obtain
  $\mathbfcal A_{mn}(\mathbf k)=\mathrm i \langle u_{m\mathbf k}(\mathbf r)|
  \bm \nabla_{\mathbf k} \mathcal H_0(\mathbf k) |u_{n\mathbf k}(\mathbf
  r)\rangle/(\varepsilon_{n\mathbf k}-\varepsilon_{m\mathbf k})$. Consequently,
  we can rewrite $\mathbf g_n(\mathbf k)=\Re\sum_{m\neq n} \langle u_{n\mathbf
  k}(\mathbf r)| \bm \nabla_{\mathbf k} \mathcal H_0(\mathbf k) |u_{m\mathbf
  k}(\mathbf r)\rangle \langle u_{m\mathbf k}(\mathbf r)| \bm \nabla_{\mathbf
  k} \mathcal H_0(\mathbf k) |u_{n\mathbf k}(\mathbf
  r)\rangle/(\varepsilon_{n\mathbf k}-\varepsilon_{m\mathbf k})^2$ and
  $\mathbfcal G_n(\mathbf k)=2\Re\sum_{m\neq n} \langle u_{n\mathbf k}(\mathbf
  r)| \bm \nabla_{\mathbf k} \mathcal H_0(\mathbf k) |u_{m\mathbf k}(\mathbf
  r)\rangle \langle u_{m\mathbf k}(\mathbf r)| \bm \nabla_{\mathbf k} \mathcal
  H_0(\mathbf k) |u_{n\mathbf k}(\mathbf r)\rangle/(\varepsilon_{n\mathbf
  k}-\varepsilon_{m\mathbf k})^3$. It is then straightforward to check that
  $\mathbfcal G_n(\mathbf k)=-\partial \mathbf g_n (\mathbf k)/\partial
  \varepsilon_{n\mathbf k}$.}
\end{thebibliography}
\end{document}

\title{Supplemental Material for\\
``A Clarification on Quantum-Metric-Induced Nonlinear Transport"}

\author{Xiao-Bin Qiang}
\email{These authors contributed equally to this work.}
\affiliation{State Key Laboratory of Quantum Functional Materials, Department of Physics, and Guangdong Basic Research Center of Excellence for Quantum Science, Southern University of Science and Technology (SUSTech), Shenzhen 518055, China}
\affiliation{Quantum Science Center of Guangdong-Hong Kong-Macao Greater Bay Area (Guangdong), Shenzhen 518045, China}

\author{Tianyu Liu}
\email{These authors contributed equally to this work.}
\affiliation{International Quantum Academy, Shenzhen 518048, China}
\affiliation{Shenzhen Key Laboratory of Quantum Science and Engineering, Shenzhen 518055, China}

\author{Zi-Xuan Gao}
\email{These authors contributed equally to this work.}
\affiliation{State Key Laboratory of Quantum Functional Materials, Department of Physics, and Guangdong Basic Research Center of Excellence for Quantum Science, Southern University of Science and Technology (SUSTech), Shenzhen 518055, China}
\affiliation{Quantum Science Center of Guangdong-Hong Kong-Macao Greater Bay Area (Guangdong), Shenzhen 518045, China}

\author{Hai-Zhou Lu}
\email{Corresponding author: luhz@sustech.edu.cn}
\affiliation{State Key Laboratory of Quantum Functional Materials, Department of Physics, and Guangdong Basic Research Center of Excellence for Quantum Science, Southern University of Science and Technology (SUSTech), Shenzhen 518055, China}
\affiliation{Quantum Science Center of Guangdong-Hong Kong-Macao Greater Bay Area (Guangdong), Shenzhen 518045, China}

\author{X. C. Xie}
\affiliation{International Center for Quantum Materials, School of Physics, Peking University, Beijing 100871, China}
\affiliation{Institute for Nanoelectronic Devices and Quantum Computing, Fudan University, Shanghai 200433, China}
\affiliation{Hefei National Laboratory, Hefei 230088, China}

\date{\today}

%\begin{abstract}
%In this Supplemental Material, we present the detailed calculations for quantum-metric-induced nonlinear conductivity by using three individual approaches: wave packet dynamics, Luttinger-Kohn approach and general pertubation theory.
%\end{abstract}

\maketitle

\tableofcontents

\section{Standard Perturbation Theory}
We now explicitly derive the electric-field-modified Berry connection and band energy using the standard perturbation theory. For a generic periodic system subject to an electric field, the Hamiltonian operator can be written as
\begin{equation} \label{H}
\hat{\mathcal{H}}=\hat{\mathcal{H}}_0+ e\mathbf{E}\cdot\hat{\mathbf{r}},
\end{equation}
where $\hat{\mathcal{H}}_0$ is the periodic Hamiltonian operator defined in the real space and $e\mathbf{E}\cdot\hat{\mathbf{r}}$ is the electric-field-induced potential with $\hat{\mathbf{r}}$ representing the position operator. For a sufficiently weak electric field $\mathbf E$, the eigenvalue problem of $\hat{\mathcal{H}}$ can be solved by perturbatively correcting the eigenenergy and eigenvector of $\hat{\mathcal{H}}_0$, respectively denoted as $\varepsilon_{n\mathbf k}$ and $|\psi_{n\mathbf k}(\mathbf r)\rangle$, where $|\psi_{n\mathbf k}(\mathbf r)\rangle$ characterizes a Bloch wave. In the context of the standard perturbation theory, the eigenvector, to the first order of $\mathbf E$, reads
\begin{equation} \label{eigvec1}
\left|\tilde \psi_{n\mathbf k} (\mathbf r) \right\rangle= |\psi_{n\mathbf k}(\mathbf r)\rangle+ \frac{V}{(2\pi)^d} \int d^d\mathbf k'\sum_{m\neq n} \frac{ \langle \psi_{m\mathbf k'}(\mathbf r) |e\mathbf E \cdot \hat{\mathbf r}| \psi_{n\mathbf k}(\mathbf r)\rangle}{\varepsilon_ {n\mathbf k}-\varepsilon_{m\mathbf k'}} |\psi_{m\mathbf k'}(\mathbf r)\rangle.
\end{equation}
Bloch's theorem requires the eigenvector of $\hat{\mathcal H}_0$ to be written as $|\psi_{n\mathbf k} (\mathbf r)\rangle = \mathrm e^{\mathrm i \mathbf k \cdot \mathbf r} |u_{n\mathbf k} (\mathbf r)\rangle$, where $|u_{n\mathbf k} (\mathbf r)\rangle$ is defined in a unit cell and has the periodicity of the Bravais lattice. The numerator in the second term in Eq.~\eqref{eigvec1} can be evaluated by using
\begin{equation} \label{polar}
\begin{split}
\langle \psi_{m\mathbf k'}(\mathbf r) | \hat{\mathbf r} |\psi_{n \mathbf k}(\mathbf r) \rangle &= \int d^d\mathbf r u_{m\mathbf k'}^*(\mathbf r)  \mathrm e^{-\mathrm i \mathbf k' \cdot \mathbf r} \mathbf r \mathrm e^{\mathrm i \mathbf k \cdot \mathbf r} u_{n \mathbf k}(\mathbf r) = -\mathrm i \int d^d\mathbf r  u_{m\mathbf k'}^*(\mathbf r)  \mathrm e^{-\mathrm i \mathbf k' \cdot \mathbf r} (\bm\nabla_{\mathbf k} \mathrm e^{\mathrm i \mathbf k \cdot \mathbf r}) u_{n \mathbf k} (\mathbf r)
\\
&=-\mathrm i \bm\nabla_{\mathbf k} \int d^d\mathbf r  u_{m\mathbf k'}^*(\mathbf r)  \mathrm e^{-\mathrm i \mathbf k' \cdot \mathbf r}  \mathrm e^{\mathrm i \mathbf k \cdot \mathbf r} u_{n \mathbf k}(\mathbf r) +  \int d^d \mathbf r \mathrm e^{\mathrm i (\mathbf k- \mathbf k') \cdot \mathbf r} u_{m\mathbf k'}^*(\mathbf r)   \mathrm i \bm\nabla_{\mathbf k}  u_{n\mathbf k}(\mathbf r)
\\
&= -\mathrm i  \bm\nabla_{\mathbf k} \langle \psi_{m\mathbf k'}(\mathbf r)  |\psi_{n \mathbf k}(\mathbf r) \rangle +  \sum_{\mathbf R} \mathrm e^{\mathrm i (\mathbf k- \mathbf k') \cdot \mathbf R} \int\limits_{\text{u.c.}} d^d \mathbf r \mathrm e^{\mathrm i (\mathbf k- \mathbf k') \cdot \mathbf r} u_{m\mathbf k'}^*(\mathbf r) \mathrm i \bm\nabla_{\mathbf k}  u_{n\mathbf k} (\mathbf r)
\\
&= \frac{(2\pi)^d}{V} \delta(\mathbf k- \mathbf k') \int\limits_{\text{u.c.}} d^d \mathbf r \mathrm e^{\mathrm i (\mathbf k- \mathbf k') \cdot \mathbf r} u_{m\mathbf k'}^*(\mathbf r) \mathrm i \bm\nabla_{\mathbf k}  u_{n\mathbf k} (\mathbf r), 
\end{split}
\end{equation}
where the integration over real space is conducted by first integrating over a unit cell and then summing over all unit cells (parameterized by the Bravais lattice vector $\mathbf R$), i.e., $\int d^d\mathbf r = \sum_{\mathbf R} \int_{\text{u.c.}} d^d\mathbf r$. When deriving Eq.~\eqref{polar}, we have used the Poisson summation formula $\sum_{\mathbf R} \mathrm e^{\mathrm i (\mathbf k- \mathbf k') \cdot \mathbf R} = \tfrac{(2\pi)^d}{V} \delta(\mathbf k-\mathbf k')$, where $\delta(\mathbf k-\mathbf k')$ is the Dirac delta function and $V$ is the volume of the unit cell. Note that the orthonormalization of the Bloch wave functions $\langle \psi_{m\mathbf k'}(\mathbf r)  |\psi_{n \mathbf k}(\mathbf r) \rangle = \tfrac{(2\pi)^d}{V}\delta_{mn}\delta(\mathbf k-\mathbf k')$ can be straightforwardly derived using the Poisson summation formula and $\langle u_{m\mathbf k}(\mathbf r)  |u_{n \mathbf k}(\mathbf r) \rangle = \delta_{mn}$, where $\delta_{mn}$ is the Kronecker delta. Plugging Eq.~\eqref{polar} into Eq.~\eqref{eigvec1}, the electric-field-modified eigenvector reads
\begin{equation} \label{eigvec2}
\begin{split}
\left|\tilde \psi_{n\mathbf k} (\mathbf r) \right\rangle &= |\psi_{n\mathbf k}(\mathbf r)\rangle+e\mathbf E \cdot \frac{V}{(2\pi)^d} \int d^d\mathbf k'\sum_{m\neq n} \frac{(2\pi)^d}{V} \delta(\mathbf k- \mathbf k') \int\limits_{\text{u.c.}} d^d \mathbf r \mathrm e^{\mathrm i (\mathbf k- \mathbf k') \cdot \mathbf r} u_{m\mathbf k'}^*(\mathbf r) \mathrm i \bm\nabla_{\mathbf k}  u_{n\mathbf k} (\mathbf r) \frac{ |\psi_{m\mathbf k'}(\mathbf r)\rangle}{\varepsilon_ {n\mathbf k}-\varepsilon_{m\mathbf k'}}
\\
&= |\psi_{n\mathbf k}(\mathbf r)\rangle+e\mathbf E \cdot  \sum_{m\neq n}   \int\limits_{\text{u.c.}} d^d \mathbf r u_{m\mathbf k}^*(\mathbf r) \mathrm i \bm\nabla_{\mathbf k}  u_{n\mathbf k} (\mathbf r) \frac{ |\psi_{m\mathbf k}(\mathbf r)\rangle}{\varepsilon_ {n\mathbf k}-\varepsilon_{m\mathbf k}}
\\
&= |\psi_{n\mathbf k}(\mathbf r)\rangle+ \sum_{m\neq n}    \frac{e\mathbf E \cdot \mathbfcal A_{mn}(\mathbf k)}{\varepsilon_ {n\mathbf k}-\varepsilon_{m\mathbf k}} |\psi_{m\mathbf k}(\mathbf r)\rangle,
\end{split}
\end{equation}
where $\mathbfcal A_{mn}(\mathbf k) = \int_{\text{u.c.}} d^d \mathbf r u_{m\mathbf k}^*(\mathbf r) \mathrm i \bm\nabla_{\mathbf k}  u_{n\mathbf k} (\mathbf r)$ is the interband Berry connection. The unit-cell-periodic part of $|\tilde \psi_{n\mathbf k} (\mathbf r) \rangle$ can be extracted as
\begin{equation}  \label{eigvec3}
|\tilde u_{n\mathbf k} (\mathbf r) \rangle= |u_{n\mathbf k}(\mathbf r)\rangle+ \sum_{m\neq n} \frac{ e\mathbf E \cdot \mathbfcal A_{mn}(\mathbf k) }{\varepsilon_ {n\mathbf k}-\varepsilon_{m\mathbf k}} |u_{m\mathbf k}(\mathbf r)\rangle.
\end{equation}
It is critically important to note that such an electric-field-modified unit-cell-periodic state has not yet been normalized.

We now calculate the electric-field-modified Berry connection for the $n$th energy band. The $n$th intraband Berry connection, to the first order of $\mathbf E$, becomes
\begin{equation} \label{Eq: An_tot}
\begin{split}
\tilde {\mathbfcal A}_n(\mathbf k) &= \frac{\langle \tilde u_{n\mathbf k} (\mathbf r) | \mathrm i\bm\nabla_{\mathbf k} |\tilde u_{n\mathbf k} (\mathbf r) \rangle}{\langle \tilde u_{n\mathbf k} (\mathbf r) | \tilde u_{n\mathbf k} (\mathbf r) \rangle} = \langle u_{n\mathbf k} (\mathbf r) | \mathrm i\bm\nabla_{\mathbf k} | u_{n\mathbf k} (\mathbf r) \rangle + \left[ \sum_{m\neq n} \frac{ e\mathbf E \cdot \mathbfcal A_{mn}(\mathbf k) }{\varepsilon_ {n\mathbf k}-\varepsilon_{m\mathbf k}} \langle u_{n\mathbf k} (\mathbf r) | \mathrm i\bm\nabla_{\mathbf k} |u_{m\mathbf k}(\mathbf r)\rangle + \text{c.c.} \right]
\\
&= \mathbfcal A_n(\mathbf k) + \left[e \sum_{m\neq n} \frac{ \mathbfcal A_{nm}(\mathbf k) \mathbfcal A_{mn}(\mathbf k) }{\varepsilon_ {n\mathbf k}-\varepsilon_{m\mathbf k}} \cdot \mathbf E + \text{c.c.} \right] = \mathbfcal A_n(\mathbf k) + 2e\Re \sum_{m\neq n} \frac{ \mathbfcal A_{nm}(\mathbf k) \mathbfcal A_{mn}(\mathbf k) }{\varepsilon_ {n\mathbf k}-\varepsilon_{m\mathbf k}} \cdot \mathbf E 
\\
&= \mathbfcal A_n(\mathbf k) + \mathbf G_n(\mathbf k) \cdot \mathbf E,
\end{split}
\end{equation}
where $\mathbf G_n(\mathbf k) = 2e\Re \sum_{m\neq n} \frac{ \mathbfcal A_{nm}(\mathbf k) \mathbfcal A_{mn}(\mathbf k) }{\varepsilon_ {n\mathbf k}-\varepsilon_{m\mathbf k}}$ is the Berry connection polarizability tensor, a gauge invariant quantity \cite{note1}. Equation~\eqref{Eq: An_tot} is identical to Eq.~(\textcolor{red}{5}) in the main text.

The evaluation of the electric-field-modified band energy is more subtle. In the context of the standard perturbation theory, the first- and second-order energy corrections respectively read
\begin{equation}
\begin{split}
\varepsilon_{n\mathbf k}^{(1)} &=e\mathbf E\cdot \langle\psi_{n\mathbf k}(\mathbf r)| \hat {\mathbf r} | \psi_{n\mathbf k}(\mathbf r) \rangle,
\\
\varepsilon_{n\mathbf k}^{(2)} &= \sum_{m\neq n}    \frac{e\mathbf E \cdot \mathbfcal A_{mn}(\mathbf k)}{\varepsilon_ {n\mathbf k}-\varepsilon_{m\mathbf k}} e\mathbf E\cdot \langle\psi_{n\mathbf k}(\mathbf r)| \hat {\mathbf r} |\psi_{m\mathbf k}(\mathbf r)\rangle.
\end{split}
\end{equation}
It is straightforward to find that both corrections diverge because $\hat{\mathbf r}$ is not a legitimate operator in a periodic system \cite{resta1998}. To circumvent the divergence, we examine the electric-field-modified unit-cell-periodic state $|\tilde u_{n\mathbf k}(\mathbf r)\rangle$. Inspired by the form of Eq.~\eqref{eigvec3}, we construct a momentum-space Hamiltonian
\begin{equation} \label{ansatz}
\mathcal H(\mathbf k)= \mathcal H_0(\mathbf k)+e\mathbf E \cdot \mathrm i \bm\nabla_{\mathbf k},
\end{equation}
where $ \mathcal H_0(\mathbf k) = \mathrm e^{-\mathrm i \mathbf k \cdot \mathbf r} \hat{\mathcal H}_0 \mathrm e^{\mathrm i \mathbf k \cdot \mathbf r}$ is the Bloch Hamiltonian with eigenenergy $\varepsilon_{n\mathbf k}$ and eigenvector $|u_{n\mathbf k}(\mathbf r)\rangle$. From the perspective of the standard perturbation theory, it is straightforward to find that the eigenvector of $\mathcal H(\mathbf k)$, to the first order of $\mathbf E$, coincides with Eq.~\eqref{eigvec3}. This suggests Eq.~\eqref{ansatz} as a valid ansatz for the periodic Hamiltonian in the momentum space \cite{nunes2001}, where the perturbation $e\mathbf E \cdot \hat{\mathbf r}$ is regularized to $e\mathbf E \cdot \mathrm i \bm\nabla_{\mathbf k}$, though $\hat{\mathbf r} \neq \mathrm e^{\mathrm i \mathbf k \cdot \mathbf r}\mathrm i \bm\nabla_{\mathbf k} \mathrm e^{-\mathrm i \mathbf k \cdot \mathbf r}$ in the mathematical sense. It is worth noting that the energy corrections become well-defined with the help of $\mathcal H(\mathbf k)$. Specifically, the first-order energy correction reads
\begin{equation}
\varepsilon_{n\mathbf k}^{(1)} =e\mathbf E\cdot \langle u_{n\mathbf k}(\mathbf r)| \mathrm i\bm\nabla_{\mathbf k} | u_{n\mathbf k}(\mathbf r) \rangle = e\mathbf E\cdot \mathbfcal A_n(\mathbf k),
\end{equation}
which converges but is not gauge invariant due to the appearance of the Berry connection $\mathbfcal A_n(\mathbf k)$. The second-order energy correction reads
\begin{equation}
\varepsilon_{n\mathbf k}^{(2)}= \sum_{m\neq n}    \frac{|\langle u_{n\mathbf k}(\mathbf r)| e\mathbf E\cdot\mathrm i\bm\nabla_{\mathbf k} | u_{m\mathbf k}(\mathbf r)\rangle|^2}{\varepsilon_ {n\mathbf k}-\varepsilon_{m\mathbf k}} = e^2\sum_{m\neq n}  \frac{\mathbf E \cdot \mathbfcal A_{nm} (\mathbf k)  \mathbfcal A_{nm}^* (\mathbf k)  \cdot \mathbf E }{\varepsilon_ {n\mathbf k}-\varepsilon_{m\mathbf k}} = \frac{e}{2} \mathbf E \cdot \mathbf G_n (\mathbf k) \cdot\mathbf E,
\end{equation}
where we have used $\mathbfcal A_{nm}^* (\mathbf k)=-\mathrm i\langle u_{n\mathbf k}(\mathbf r)|\bm\nabla_{\mathbf k} u_{m\mathbf k}(\mathbf r) \rangle^* = -\mathrm i \langle \bm\nabla_{\mathbf k} u_{m\mathbf k}(\mathbf r) | u_{n\mathbf k}(\mathbf r) \rangle = \langle u_{m\mathbf k}(\mathbf r) |\mathrm i \bm \nabla_{\mathbf k}| u_{n\mathbf k}(\mathbf r) \rangle = \mathbfcal A_{mn}(\mathbf k)$ and  $|\mathbf E\cdot \mathbfcal A_{mn}(\mathbf k)|^2=\mathbf E\cdot \Re[\mathbfcal A_{nm}(\mathbf k) \mathbfcal A_{mn}(\mathbf k)] \cdot\mathbf E$. Unlike $\varepsilon_{n\mathbf k}^{(1)}$, $\varepsilon_{n\mathbf k}^{(2)}$ possesses gauge invariance, which is inherited from the Berry connection polarizability tensor \cite{note1}. Consequently, the electric-field-modified band energy should be written as
\begin{equation} \label{Eq: en_tot}
\tilde \varepsilon_{n\mathbf k} = \varepsilon_{n\mathbf k} + \varepsilon_{n\mathbf k}^{(2)} = \varepsilon_{n\mathbf k} + \frac{e}{2} \mathbf E \cdot \mathbf G_n (\mathbf k) \cdot\mathbf E,
\end{equation}
which is gauge invariant and observable. Equation~\eqref{Eq: en_tot} is identical to Eq.~(\textcolor{red}{7}) in the main text.

\section{Wave Packet Dynamics}
\label{sec2}
We now derive the electric-field-modified Berry connection and band energy in the framework of the wave packet dynamics. We will start with a general formalism on wave packet construction and then investigate the effect of the electric field on the Berry connection and band energy.

\subsection{Wave Packet Construction\label{Sec: WP2_E}}
\label{sec2a}
For a wave packet centered at $\mathbf r_c$ with a narrow spatial spread, it would be reasonable to rearrange the model Hamiltonian to
\begin{equation}
\hat{\mathcal H}=\hat{\mathcal H}_c+e\mathbf E \cdot (\hat{\mathbf r}-\mathbf r_c),
\end{equation}
where $\hat{\mathcal H}_c=\hat{\mathcal H}_0+e\mathbf E\cdot\mathbf r_c$ is the local Hamiltonian ``felt'' by the wave packet and $e\mathbf E \cdot (\hat{\mathbf r}-\mathbf r_c)$ is the electric-field-induced perturbation to the wave packet. As $e\mathbf E\cdot\mathbf r_c$ is a constant, $\hat{\mathcal H}_c$ and $\hat{\mathcal H}_0$ exhibit the same periodicity and share the same Bloch eigenvectors. Explicitly, the eigenvalue equation of $\hat{\mathcal H}_c$ reads
\begin{equation}
\hat{\mathcal{H}}_c|\psi_{n\mathbf{k}}(\mathbf r)\rangle=\varepsilon_{n\mathbf k}^c|\psi_{n\mathbf{k}}(\mathbf r)\rangle,
\end{equation}
where $\varepsilon_{n\mathbf{k}}^c=\varepsilon_{n \mathbf{k}}+e\mathbf E\cdot\mathbf r_c$ is the $n$th eigenenergy. The wave packet can be conveniently constructed in terms of the Bloch eigenvectors as
\begin{equation} \label{wp}
|W_n(\mathbf r)\rangle= \int d^d\mathbf{k} w_n(\mathbf{k}) |\psi_{n\mathbf{k}}(\mathbf r)\rangle,
\end{equation}
where $w_n(\mathbf k)$ represents the superposition weight. The normalization of the wave packet requires
\begin{equation}
\begin{split}
\langle W_n(\mathbf r)| W_n(\mathbf r) \rangle &= \int d^d\mathbf k \int d^d\mathbf k' w_n^*(\mathbf k) w_n(\mathbf k') \langle \psi_{n\mathbf k}(\mathbf r)| \psi_{n\mathbf k'}(\mathbf r) \rangle = \frac{(2\pi)^d}{V} \int d^d\mathbf k \int d^d\mathbf k' w_n^*(\mathbf k) w_n(\mathbf k')\delta(\mathbf k-\mathbf k')
\\
&= \frac{(2\pi)^d}{V} \int d^d\mathbf k |w_n(\mathbf k)|^2 = 1,
\end{split}
\end{equation}
which suggests that the superposition weight satifies $|w_n(\mathbf k)|^2=\tfrac{V}{(2\pi)^d}\delta(\mathbf{k}-\mathbf{k}_c)$, with $\mathbf{k}_c$ labeling the momentum center of the wave packet.

In the presence of the electric-field-induced perturbation, the wave packet can develop mixing with other bands labeled by $m\neq n$ and is thus modified to
\begin{equation}\label{Eq: WP2_E}
\ket{\tilde{W}_n(\mathbf r)}=\int d^d\mathbf{k} \left[ (1-\delta)w_n(\mathbf{k}) | \psi_{n\mathbf{k}} (\mathbf r) \rangle+\sum_{m\neq n} w_m^{(1)}(\mathbf{k})| \psi_{m\mathbf{k}} (\mathbf r) \rangle\right],    
\end{equation}
where the parameter $\delta$ and the coefficient $w_m^{(1)}(\mathbf k)$ quantify the interband mixing. The normalization of the modified wave packet requires $\delta=\tfrac{1}{2}\tfrac{(2\pi)^d}{V}\int d^d\mathbf k \sum_{m\neq n}|w_m^{(1)}(\mathbf k)|^2$. To determine $w_m^{(1)}(\mathbf k)$, we require $|\tilde W_n(\mathbf r)\rangle$ to be an eigenvector of $\hat{\mathcal H}$. We then multiply $\langle \psi_{l\neq n \mathbf k}(\mathbf r) |$ to the eigenvalue equation and get
\begin{equation} \label{eigen}
\Big\langle \psi_{l\mathbf k}(\mathbf r) \Big|\hat{\mathcal H} \Big|\tilde W_n(\mathbf r)\Big\rangle = \tilde \varepsilon_{n\mathbf k} \Big\langle \psi_{l\mathbf k}(\mathbf r) \Big|\tilde W_n(\mathbf r) \Big\rangle.
\end{equation}
The left-hand side of Eq.~\eqref{eigen}, to the first order of $\mathbf E$, reads
\begin{equation} \label{lhs}
\begin{aligned}
\Big\langle \psi_{l\mathbf k}(\mathbf r)\Big| \hat{\mathcal{H}}\Big|\tilde{W}_n(\mathbf r)\Big\rangle &=\int d^d\mathbf{k}'\left[ \langle \psi_{l\mathbf k}(\mathbf r)| (\hat{\mathcal H}_0+e\mathbf{E}\cdot\hat{\mathbf{r}})(1-\delta)w_n (\mathbf k') | \psi_{n\mathbf k'}(\mathbf r) \rangle+\sum_{m\neq n} \langle \psi_{l \mathbf k}(\mathbf r)| (\hat{\mathcal H}_0+e\mathbf{E}\cdot\hat{\mathbf{r}})w_m^{(1)}(\mathbf k')|\psi_{m\mathbf k'}(\mathbf r)\rangle\right]
\\
& \simeq\int d^d\mathbf{k}'  w_n(\mathbf k') e\mathbf{E}\cdot \langle \psi_{l\mathbf k}(\mathbf r) | \hat{\mathbf r} |\psi_{n \mathbf k'}(\mathbf r)\rangle+ \int d^d\mathbf{k}' \sum_{m\neq n} \varepsilon_{l\mathbf k} w_m^{(1)}(\mathbf k') \langle \psi_{l\mathbf k}(\mathbf r) |\psi_{m \mathbf k'}(\mathbf r)\rangle
\\
&=\frac{(2\pi)^d}{V} \int d^d\mathbf{k}'  \delta(\mathbf k- \mathbf k') w_n(\mathbf k') e\mathbf{E}\cdot    \int\limits_{\text{u.c.}} d^d\mathbf r \mathrm e^{\mathrm i (\mathbf k'- \mathbf k) \cdot \mathbf r} u_{l\mathbf k}^*(\mathbf r)  \mathrm i \bm\nabla_{\mathbf k'}  u_{n\mathbf k'} (\mathbf r) + \frac{(2\pi)^d}{V} \varepsilon_{l\mathbf k} w_l^{(1)}(\mathbf k)
\\
&= \frac{(2\pi)^d}{V}   w_n(\mathbf k) e\mathbf{E}\cdot    \int\limits_{\text{u.c.}} d^d\mathbf r u_{l\mathbf k}^*(\mathbf r)  \mathrm i \bm\nabla_{\mathbf k}  u_{n\mathbf k} (\mathbf r) + \frac{(2\pi)^d}{V} \varepsilon_{l\mathbf k} w_l^{(1)}(\mathbf k)
\\
&= \frac{(2\pi)^d}{V}   \left[ w_n(\mathbf k) e\mathbf{E}\cdot \mathbfcal A_{ln}(\mathbf k) + \varepsilon_{l\mathbf k} w_l^{(1)}(\mathbf k)  \right],
\end{aligned}
\end{equation}
where we have used Eq.~\eqref{polar} and $\langle \psi_{m\mathbf k'}(\mathbf r)  |\psi_{n \mathbf k}(\mathbf r) \rangle = \tfrac{(2\pi)^d}{V}\delta_{mn}\delta(\mathbf k-\mathbf k')$. In the meanwhile, the right-hand side of Eq.~\eqref{eigen}, to the first order of $\mathbf E$, reads
\begin{equation} \label{rhs}
\begin{aligned}
\tilde{\varepsilon}_{n\mathbf k} \Big \langle  \psi_{l\mathbf{k}}(\mathbf r) \Big| \tilde{W}_n(\mathbf r) \Big \rangle
&=\tilde{\varepsilon}_{n\mathbf k}  \int d^d\mathbf{k}' \left[  \langle  \psi_{l\mathbf{k}}(\mathbf r)| \psi_{n\mathbf{k}'}(\mathbf r)\rangle (1-\delta) w_n (\mathbf{k}')+ \sum_{m\neq n}  \langle \psi_{l\mathbf{k}}(\mathbf r)|\psi_{m \mathbf{k}'}(\mathbf r)\rangle w_m^{(1)} (\mathbf{k}') \right]
\\
&= \frac{(2\pi)^d}{V} \tilde{\varepsilon}_{n\mathbf k} \int d^d\mathbf{k}'  \delta(\mathbf{k}-\mathbf{k}')  \sum_{m\neq n} \delta_{lm}w_m^{(1)} (\mathbf{k}')
\\
&\simeq \frac{(2\pi)^d}{V} \varepsilon_{n\mathbf k} w_l^{(1)} (\mathbf{k}).
\end{aligned}
\end{equation}
Making use of Eqs.~\eqref{lhs} and~\eqref{rhs}, it is straightforward to find that
\begin{equation}
w_l^{(1)} (\mathbf{k})=\frac{e\mathbf{E}\cdot \mathbfcal A_{ln}(\mathbf k)}{\varepsilon_{n\mathbf k} - \varepsilon_{l\mathbf k} } w_n(\mathbf k). \end{equation}
The wave packet can then be explicitly constructed as
\begin{equation}\label{Eq: WP_E}
\ket{\tilde{W}_n(\mathbf r)}=\int d^d\mathbf{k}\left[(1-\delta)w_n(\mathbf{k})| \psi_{n\mathbf{k}}(\mathbf r)\rangle+\sum_{m\neq n} \mathcal{C}_{mn}(\mathbf k)w_n(\mathbf{k})|\psi_{m\mathbf{k}}(\mathbf r)\rangle\right],
\end{equation}
where we have defined the coefficient $\mathcal{C}_{mn}(\mathbf k)=\frac{ e\mathbf{E}\cdot\mathbfcal{A}_{mn}(\mathbf k)}{\varepsilon_{n\mathbf k}-\varepsilon_{m\mathbf k}}$. We note that matrix $\mathcal{C}(\mathbf k)$ is anti-Hermitian as $\mathcal{C}_{mn}(\mathbf k)^*=\frac{ e\mathbf{E}\cdot\mathbfcal{A}_{mn}^*(\mathbf k)}{\varepsilon_{n\mathbf k}-\varepsilon_{m\mathbf k}}=-\frac{ e\mathbf{E}\cdot\mathbfcal{A}_{nm}(\mathbf k)}{\varepsilon_{m\mathbf k}-\varepsilon_{n\mathbf k}}=-\mathcal C_{nm}(\mathbf k)$. To the first order of $\mathbf E$, the wave packet becomes unnormalized and reads
\begin{equation}
\ket{\tilde{W}_n(\mathbf r)}\simeq \int d^d\mathbf{k} w_n(\mathbf{k}) \left[| \psi_{n\mathbf{k}}(\mathbf r)\rangle+\sum_{m\neq n} \frac{ e\mathbf{E}\cdot\mathbfcal{A}_{mn}(\mathbf k)}{\varepsilon_{n\mathbf k}-\varepsilon_{m\mathbf k}}|\psi_{m\mathbf{k}}(\mathbf r)\rangle\right] = \int d^d\mathbf{k} w_n(\mathbf{k}) \left| \tilde \psi_{n\mathbf k} (\mathbf r)\right\rangle,
\end{equation}
which is an analog to the field-free wave packet [Eq.~\eqref{wp}] with the electric-field-modified Bloch state $| \tilde \psi_{n\mathbf k} (\mathbf r)\rangle$ [Eq.~\eqref{eigvec2}] taking over the role of the Bloch eigenvector $| \psi_{n\mathbf k} (\mathbf r)\rangle$.

\subsection{Positional Shift}
\label{sec2b}
Because of the interband mixing, the real-space center of the wave packet should also shift accordingly. The field-free wave packet is centered at
\begin{equation} \label{rc}
\begin{split}
\mathbf r_c =\langle W_n(\mathbf r) | \hat {\mathbf r} | W_n(\mathbf r) \rangle =& \int d^d\mathbf k \int d^d \mathbf k' w_n^*(\mathbf k') w_n(\mathbf k) \langle \psi_{n\mathbf k'} (\mathbf r) | \hat{\mathbf r} | \psi_{n\mathbf k} (\mathbf r) \rangle
\\
=& -\mathrm i \int d^d\mathbf k \int d^d \mathbf k' w_n^*(\mathbf k') w_n(\mathbf k)  \bm\nabla_{\mathbf k} \langle \psi_{n\mathbf k'} (\mathbf r) | \psi_{n\mathbf k} (\mathbf r) \rangle 
\\
&+ \frac{(2\pi)^d}{V} \int d^d\mathbf k \int d^d \mathbf k' w_n^*(\mathbf k') w_n(\mathbf k) \delta(\mathbf k -\mathbf k') \int\limits_{\text{u.c.}} d^d\mathbf r \mathrm e^{\mathrm i (\mathbf k-\mathbf k')\cdot\mathbf r} u_{n\mathbf k'}^*(\mathbf r) \mathrm i\bm \nabla_{\mathbf k} u_{n\mathbf k}(\mathbf r),
\end{split}
\end{equation}
where we have used Eq.~\eqref{polar}. The first term in Eq.~\eqref{rc} can be further simplified by integrating by parts as
\begin{equation}
\begin{split}
-&\mathrm i \int d^d\mathbf k \int d^d \mathbf k' w_n^*(\mathbf k') w_n(\mathbf k)  \bm\nabla_{\mathbf k} \langle \psi_{n\mathbf k'} (\mathbf r) | \psi_{n\mathbf k} (\mathbf r) \rangle = \mathrm i \int d^d\mathbf k \int d^d \mathbf k' w_n^*(\mathbf k')   [\bm\nabla_{\mathbf k}w_n(\mathbf k)] \langle \psi_{n\mathbf k'} (\mathbf r) | \psi_{n\mathbf k} (\mathbf r) \rangle
\\
&= \frac{(2\pi)^d}{V} \mathrm i \int d^d\mathbf k \int d^d \mathbf k' w_n^*(\mathbf k')   [\bm\nabla_{\mathbf k}w_n(\mathbf k)] \delta(\mathbf k-\mathbf k') = \frac{(2\pi)^d}{V} \mathrm i \int d^d\mathbf k w_n^*(\mathbf k)   [\bm\nabla_{\mathbf k}w_n(\mathbf k)] 
\\
& = \frac{(2\pi)^d}{V} \mathrm i \int d^d\mathbf k |w_n(\mathbf k)| \mathrm e^{-\mathrm i\arg w_n(\mathbf k)}   \left[\mathrm e^{\mathrm i\arg w_n(\mathbf k)}\bm\nabla_{\mathbf k}|w_n(\mathbf k)| +\mathrm i \bm\nabla_{\mathbf k}  \arg w_n(\mathbf k) \mathrm e^{\mathrm i\arg w_n(\mathbf k)} |w_n(\mathbf k)| \right]
\\
& = -\frac{(2\pi)^d}{V} \int d^d\mathbf k |w_n(\mathbf k)|^2 \bm\nabla_{\mathbf k}  \arg w_n(\mathbf k) = -\int d^d\mathbf k \delta(\mathbf k-\mathbf k_c) \bm\nabla_{\mathbf k}  \arg w_n(\mathbf k)
\\
& = - \bm\nabla_{\mathbf k}  \arg w_n(\mathbf k) |_{\mathbf k=\mathbf k_c},
\end{split}
\end{equation}
where we resort to the facts that $\int d^d\mathbf k \bm \nabla_{\mathbf k} [w_n(\mathbf k) \langle \psi_{n\mathbf k'} (\mathbf r) | \psi_{n\mathbf k} (\mathbf r) \rangle]=0$ and $\int d^d\mathbf k \bm \nabla_{\mathbf k} |w_n(\mathbf k)|^2=0$. In the meanwhile, the second term in Eq.~\eqref{rc} reads
\begin{equation}
\begin{split}
&\frac{(2\pi)^d}{V} \int d^d\mathbf k \int d^d \mathbf k' w_n^*(\mathbf k') w_n(\mathbf k) \delta(\mathbf k -\mathbf k') \int\limits_{\text{u.c.}} d^d\mathbf r \mathrm e^{\mathrm i (\mathbf k-\mathbf k')\cdot\mathbf r} u_{n\bm k'}^*(\mathbf r) \mathrm i\bm \nabla_{\mathbf k} u_{n\mathbf k}(\mathbf r)
\\
&= \frac{(2\pi)^d}{V} \int d^d\mathbf k |w_n(\mathbf k)|^2 \int\limits_{\text{u.c.}} d^d\mathbf r  u_{n\bm k}^*(\mathbf r) \mathrm i\bm \nabla_{\mathbf k} u_{n\mathbf k}(\mathbf r) =  \int d^d\mathbf k \delta(\mathbf k-\mathbf k_c) \mathbfcal A_n(\mathbf k)
\\
& = \mathbfcal A_n(\mathbf k_c).
\end{split}
\end{equation}
Therefore, the real-space center of the field-free wave packet is
\begin{equation} \label{rc2}
\mathbf r_c = - \bm\nabla_{\mathbf k}  \arg w_n(\mathbf k) |_{\mathbf k=\mathbf k_c} +  \mathbfcal A_n(\mathbf k_c).
\end{equation}
It is worth noting that $\mathbf r_c$ depends on gauge. On the one hand, although $|w_n(\mathbf k)|^2=\tfrac{V}{(2\pi)^d}\delta(\mathbf k-\mathbf k_c)$, there is no requirement on $\arg w_n(\mathbf k)$, which can be chosen arbitrarily. On the other hand, the Berry connection also depends on gauge. 

In the presence of the applied electric field, the real-space center of the wave packet should in general acquire a shift. The new center is located, to the first order of $\mathbf E$, at
\begin{align} \label{trc}
\tilde{\mathbf r}_c &= \left \langle \tilde{W}_n(\mathbf r) \right| \hat{\mathbf{r}} \left |\tilde{W}_n(\mathbf r)\right \rangle \simeq \langle W_n(\mathbf r) |\hat{\mathbf r} |W_n(\mathbf r)\rangle + \left[\int d^d \mathbf k \sum_{m\neq n}\frac{e\mathbf E\cdot \mathbfcal A_{mn}(\mathbf k)}{\varepsilon_{n\mathbf k}-\varepsilon_{m\mathbf k}} w_n(\mathbf k) \langle W_n(\mathbf r)| \mathbf r |\psi_{m\mathbf k}(\mathbf r) \rangle +\text{c.c.}\right] \nonumber
\\
&=\mathbf r_c + \left[\int d^d \mathbf k \int d^d \mathbf k' \sum_{m\neq n} \frac{e\mathbf E\cdot \mathbfcal A_{mn}(\mathbf k)}{\varepsilon_{n\mathbf k}-\varepsilon_{m\mathbf k}} w_n^*(\mathbf k') w_n(\mathbf k) \langle \psi_{n\mathbf k'}(\mathbf r)| \mathbf r |\psi_{m\mathbf k}(\mathbf r) \rangle +\text{c.c.}\right] \nonumber
\\
&=\mathbf r_c +\frac{(2\pi)^d}{V} \left[\int d^d \mathbf k \int d^d \mathbf k' \sum_{m\neq n}\frac{e\mathbf E\cdot \mathbfcal A_{mn}(\mathbf k)}{\varepsilon_{n\mathbf k}-\varepsilon_{m\mathbf k}} w_n^*(\mathbf k') w_n(\mathbf k) \delta(\mathbf k-\mathbf k') \int\limits_{\text{u.c.}} d^d\mathbf r \mathrm e^{\mathrm i (\mathbf k- \mathbf k') \cdot \mathbf r} u_{n\mathbf k'}^*(\mathbf r)  \mathrm i \bm\nabla_{\mathbf k}  u_{m\mathbf k} (\mathbf r) +\text{c.c.}\right] \nonumber
\\
&=\mathbf r_c +\frac{(2\pi)^d}{V} \left[\int d^d \mathbf k  \sum_{m\neq n} \frac{e\mathbf E\cdot \mathbfcal A_{mn}(\mathbf k)}{\varepsilon_{n\mathbf k}-\varepsilon_{m\mathbf k}}  w_n^*(\mathbf k) w_n(\mathbf k)  \int\limits_{\text{u.c.}} d\mathbf r u_{n\mathbf k}^*(\mathbf r)  \mathrm i \bm\nabla_{\mathbf k}  u_{m\mathbf k} (\mathbf r) +\text{c.c.}\right] \nonumber
\\
&=\mathbf r_c + \left[\int d^d \mathbf k  \sum_{m\neq n} \frac{e\mathbf E\cdot \mathbfcal A_{mn}(\mathbf k)}{\varepsilon_{n\mathbf k}-\varepsilon_{m\mathbf k}} \delta(\mathbf k-\mathbf k_c)  \mathbfcal A_{nm}(\mathbf k) +\text{c.c.}\right] = \mathbf r_c + \left[\int d^d \mathbf k  \sum_{m\neq n}\frac{e \mathbfcal A_{nm}(\mathbf k) \mathbfcal A_{mn}(\mathbf k) \cdot \mathbf E}{\varepsilon_{n\mathbf k}-\varepsilon_{m\mathbf k}} \delta(\mathbf k-\mathbf k_c)  +\text{c.c.}\right] \nonumber
\\
&= \mathbf r_c +  \left[ 2e\Re \sum_{m\neq n}\frac{\mathbfcal A_{nm}(\mathbf k_c) \mathbfcal A_{mn}(\mathbf k_c)}{\varepsilon_{n\mathbf k_c}-\varepsilon_{m\mathbf k_c}} \right] \cdot \mathbf E = \mathbf r_c + \mathbf G_n(\mathbf k_c) \cdot \mathbf E \nonumber
\\
&= - \bm\nabla_{\mathbf k}  \arg w_n(\mathbf k) |_{\mathbf k=\mathbf k_c} +  \mathbfcal A_n(\mathbf k_c) + \mathbf G_n(\mathbf k_c) \cdot \mathbf E.
\end{align}
According to Eq.~\eqref{trc}, it is straightforward to find that the applied electric field shifts the real-space center of the wave packet by $\tilde{\mathbf r}_c- \mathbf r_c=\mathbf G_n (\mathbf k_c) \cdot \mathbf E$. Although both $\tilde{\mathbf r}_c$ and $\mathbf r_c$ depend on gauge, the positional shift itself is indeed gauge invariant because of the appearance of Berry connection polarizability \cite{note1}. We choose a specific gauge by performing $|u_{n\mathbf k}(\mathbf r)\rangle \rightarrow \mathrm e^{-\mathrm i \arg w_n(\mathbf k)} |u_{n\mathbf k}(\mathbf r)\rangle$. The Berry connection is then transformed according to $\mathbfcal A_n(\mathbf k)\rightarrow \mathbfcal A_n(\mathbf k) + \bm\nabla_{\mathbf k} \arg w_n(\mathbf k)$, and the real-space centers are located at $\mathbf r_c= \mathbfcal A_n(\mathbf k_c)$ and $\tilde{\mathbf r}_c= \mathbfcal A_n(\mathbf k_c)+ \mathbf G_n(\mathbf k_c) \cdot \mathbf E$. Therefore, the positional shift can be treated as a correction to the Berry connection. The electric-field-modified Berry connection then reads
\begin{equation}
\tilde {\mathbfcal A}_n(\mathbf k_c)= \mathbfcal A_n(\mathbf k_c) + \mathbf G_n(\mathbf k_c) \cdot \mathbf E.
\end{equation}
This expression is consistent with Eq.~\eqref{Eq: An_tot} except that the argument here is the momentum center of the wave packet $\mathbf k_c=\int d^d \mathbf k |w_n(\mathbf k)|^2\mathbf k$ rather than the crystal momentum $\mathbf k$.

\subsection{Wave Packet Energy}
\label{sec2c}
We now calculate the energy of the wave packet. The contribution from the local Hamiltonian, to the second order of $\mathbf E$, reads
\begin{align}
\left \langle \tilde{W}_n(\mathbf r) \right|\hat{\mathcal{H}}_c\left|\tilde{W}_n(\mathbf r)\right\rangle=& \int d^d\mathbf k \int d^d\mathbf k' (1-\delta)^2 w_n^*(\mathbf k') w_n(\mathbf k) \left\langle \psi_{n\mathbf k'}(\mathbf r)\left| \hat{\mathcal{H}}_c\right| \psi_{n\mathbf k}(\mathbf r) \right\rangle \nonumber
\\
&+ \int d^d\mathbf k \int d^d\mathbf k' \sum_{l\neq n}\sum_{m\neq n} [w_l^{(1)}(\mathbf k')]^* w_m^{(1)}(\mathbf k) \left\langle \psi_{l\mathbf k'}(\mathbf r)\left| \hat{\mathcal{H}}_c\right| \psi_{m\mathbf k}(\mathbf r) \right\rangle \nonumber
\\
= &\frac{(2\pi)^d}{V} \int d^d\mathbf k \int d^d\mathbf k' \delta(\mathbf k-\mathbf k') \left[(1-\delta)^2 w_n^*(\mathbf k') w_n(\mathbf k) \varepsilon_{n\mathbf k}^c +\sum_{l\neq n}\sum_{m\neq n} [w_l^{(1)}(\mathbf k')]^* w_m^{(1)}(\mathbf k) \delta_{lm}\varepsilon_{m\mathbf k}^c\right] \nonumber
\\
=& \frac{(2\pi)^d}{V} \int d^d\mathbf k \left[(1-\delta)^2 |w_n(\mathbf k)|^2 \varepsilon_{n\mathbf k}^c +\sum_{m\neq n} |w_m^{(1)}(\mathbf k)|^2 \varepsilon_{m\mathbf k}^c\right] \nonumber
\\
=& (1-\delta)^2 \int d^d\mathbf k  \delta(\mathbf k-\mathbf k_c) \varepsilon_{n\mathbf k}^c +  \frac{(2\pi)^d}{V} \int d^d\mathbf k \sum_{m\neq n} |w_m^{(1)}(\mathbf k)|^2 \varepsilon_{m\mathbf k}^c \nonumber
\\
=&\left[ 1- \frac{(2\pi)^d}{V} \int d^d\mathbf k \sum_{m\neq n} |w_m^{(1)}(\mathbf k)|^2 \right] \varepsilon_{n\mathbf k_c}^c + \frac{(2\pi)^d}{V} \int d^d\mathbf k \sum_{m\neq n} |w_m^{(1)}(\mathbf k)|^2 \varepsilon_{m\mathbf k}^c \nonumber
\\
= & \varepsilon_{n\mathbf k_c}^c - \frac{(2\pi)^d}{V} \int d^d\mathbf k \sum_{m\neq n} \left|\frac{e\mathbf E\cdot \mathbfcal A_{mn}(\mathbf k)}{\varepsilon_{n\mathbf k}-\varepsilon_{m\mathbf k}}\right|^2 |w_n(\mathbf k)|^2 (\varepsilon_{n\mathbf k_c}^c - \varepsilon_{m\mathbf k}^c) \nonumber
\\
=& \varepsilon_{n\mathbf k_c}^c - \int d^d\mathbf k \delta(\mathbf k-\mathbf k_c) \sum_{m\neq n} \left|\frac{e\mathbf E\cdot \mathbfcal A_{mn}(\mathbf k)}{\varepsilon_{n\mathbf k}-\varepsilon_{m\mathbf k}}\right|^2  (\varepsilon_{n\mathbf k_c}^c - \varepsilon_{m\mathbf k}^c) \nonumber
\\
=& \varepsilon_{n\mathbf k_c}^c- \sum_{m\neq n} \left|\frac{e\mathbf E\cdot \mathbfcal A_{mn}(\mathbf k_c)}{\varepsilon_{n\mathbf k_c}-\varepsilon_{m\mathbf k_c}}\right|^2  (\varepsilon_{n\mathbf k_c}^c - \varepsilon_{m\mathbf k_c}^c) \simeq  \varepsilon_{n\mathbf k_c}^c - e^2\sum_{m\neq n} \frac{|\mathbf E\cdot \mathbfcal A_{mn}(\mathbf k_c)|^2}{\varepsilon_{n\mathbf k_c}-\varepsilon_{m\mathbf k_c}} \nonumber
\\
=& \varepsilon_{n\mathbf k_c} + e\mathbf E\cdot\mathbf r_c - e^2 \mathbf E\cdot \sum_{m\neq n} \frac{ \Re[\mathbfcal A_{nm}(\mathbf k_c) \mathbfcal A_{mn}(\mathbf k_c)]}{\varepsilon_{n\mathbf k_c}-\varepsilon_{m\mathbf k_c}} \cdot \mathbf E \nonumber
\\
=& \varepsilon_{n\mathbf k_c} + e\mathbf E\cdot\mathbf r_c - \frac{e}{2} \mathbf E \cdot \mathbf G_n(\mathbf k_c) \cdot \mathbf E,
\end{align}
where we notice the fact that $|\mathbf E\cdot \mathbfcal A_{mn}(\mathbf k_c)|^2 =\mathbf E\cdot \Re[\mathbfcal A_{nm}(\mathbf k_c) \mathbfcal A_{mn}(\mathbf k_c)] \cdot\mathbf E$. In the meanwhile, the contribution from the perturbation reads
\begin{equation}
\begin{aligned}
\left \langle \tilde{W}_n(\mathbf r) \right| \mathbf E\cdot(\hat{\mathbf r}-\mathbf r_c) \left|\tilde{W}_n(\mathbf r)\right\rangle =e\mathbf{E}\cdot \left[ \left \langle \tilde{W}_n(\mathbf r) \right| \hat{\mathbf r} \left|\tilde{W}_n(\mathbf r) \right\rangle -\mathbf r_c \right] = e\mathbf{E}\cdot \mathbf G_n(\mathbf k_c) \cdot \mathbf{E},
\end{aligned}
\end{equation}
where Eq.~\eqref{trc} is used. The total energy of the wave packet in the presence of the electric field then becomes
\begin{equation}
\left \langle \tilde{W}_n(\mathbf r) \left|\hat{\mathcal{H}}\right|\tilde{W}_n(\mathbf r)\right\rangle=\left \langle \tilde{W}_n(\mathbf r) \right| \left[\hat{\mathcal{H}}_c +  \mathbf E\cdot(\hat{\mathbf r}-\mathbf r_c) \right] \left|\tilde{W}_n(\mathbf r)\right\rangle = \varepsilon_{n\mathbf k_c} + e\mathbf E\cdot\mathbf r_c + \frac{e}{2} \mathbf E \cdot \mathbf G_n(\mathbf k_c) \cdot \mathbf E,
\end{equation}
where the second term is not observable because $\mathbf r_c = - \bm\nabla_{\mathbf k}  \arg w_n(\mathbf k) |_{\mathbf k=\mathbf k_c} +  \mathbfcal A_n(\mathbf k_c)$ depends on gauge (see Sec.~\ref{sec2b}). In fact, this term can be regarded as the electric dipole correction to the wave packet \cite{GaoY19fop}. The wave packet energy should be written as
\begin{equation}
\tilde \varepsilon_{n\mathbf k_c}= \varepsilon_{n\mathbf k_c} + \frac{e}{2} \mathbf E \cdot \mathbf G_n(\mathbf k_c) \cdot \mathbf E,
\end{equation}
which is consistent with the electric-field-modified band energy derived from the standard perturbation theory [Eq.~\eqref{Eq: en_tot}] except that the argument here is $\mathbf k_c$ rather than $\mathbf k$.

\section{Luttinger-Kohn Approach}
\label{sec3}
To extract the electric-field-modified Berry connection and band energy, we now consider a different formalism known as the Luttinger-Kohn approach \cite{Luttinger55pr}. To implement the approach, we first rewrite the ansatz [Eq.~\eqref{ansatz}] as
\begin{equation}
\mathcal H(\mathbf k)=\mathcal H_0(\mathbf k)+\lambda e\mathbf E\cdot\mathrm i\bm\nabla_{\mathbf k},
\end{equation}
where the parameter $\lambda$ is introduced to track the order of perturbation and can be conveniently set to unity as needed. We then perform a unitary Schrieffer-Wolff transformation $|u_{n\mathbf k}(\mathbf r)\rangle \rightarrow \mathrm e^{\lambda\mathcal S(\mathbf k)} |u_{n\mathbf k}(\mathbf r)\rangle$, where the generator $\mathcal S(\mathbf k)$ is anti-Hermitian to guarantee the unitarity of the transformation \cite{Schrieffer66pr}, such that $\mathrm e^{\lambda\mathcal S(\mathbf k)} |u_{n\mathbf k}(\mathbf r)\rangle$ diagonalizes $\mathcal H(\mathbf k)$ to the desired order of $\lambda$ (or, equivalently, $\mathbf E$) with an appropriate choice of the generator $\mathcal S(\mathbf k)$. The unitarily transformed Hamiltonian reads
\begin{equation} \label{Heff}
\begin{split}
\mathcal{H}_{\text{eff}}(\mathbf k) &=\mathrm e^{-\lambda \mathcal{S}(\mathbf k)}\mathcal{H}(\mathbf k) \mathrm e^{\lambda \mathcal{S}(\mathbf k)} =\mathcal{H}(\mathbf k)+\lambda[\mathcal{H}(\mathbf k),\mathcal{S}(\mathbf k)]+\frac{\lambda^2}{2}[[\mathcal{H}(\mathbf k),\mathcal S(\mathbf k)],\mathcal{S}(\mathbf k)]+\cdots
\\
&=\mathcal{H}_0(\mathbf k)+\lambda e\mathbf E\cdot\mathrm i\bm\nabla_{\mathbf k}+\lambda[\mathcal{H}_0(\mathbf k)+\lambda e\mathbf E\cdot\mathrm i\bm\nabla_{\mathbf k}, \mathcal{S}(\mathbf k)]+\frac{\lambda^2}{2}[[\mathcal{H}_0(\mathbf k)+\lambda e\mathbf E\cdot\mathrm i\bm\nabla_{\mathbf k}, \mathcal{S}(\mathbf k)],\mathcal{S}(\mathbf k)]+\cdots
\\
&=\mathcal{H}_0(\mathbf k)+\lambda\big(e\mathbf E\cdot\mathrm i\bm\nabla_{\mathbf k}+ [\mathcal{H}_0(\mathbf k), \mathcal{S}(\mathbf k)]\big)+ \lambda^2\left([e\mathbf E\cdot\mathrm i\bm\nabla_{\mathbf k}, \mathcal{S}(\mathbf k)]+\frac{1}{2}[[\mathcal{H}_0(\mathbf k),\mathcal S(\mathbf k)],\mathcal{S}(\mathbf k)]  \right)+\mathcal O(\lambda^3),
\end{split}
\end{equation}
where the Baker-Campbell-Hausdorff formula is used. Here, we would expect $\mathrm e^{\lambda\mathcal S(\mathbf k)} |u_{n\mathbf k}(\mathbf r)\rangle$ to diagonalize $\mathcal H(\mathbf k)$ to the first order of $\lambda$. This is equivalent to that $|u_{n\mathbf k}(\mathbf r)\rangle$ diagonalizes $\mathcal{H}_{\text{eff}}(\mathbf k)$ to the first order of $\lambda$ such that the energy correction arising from the third term in Eq.~\eqref{Heff} (scaled as $\lambda^2$) coincides with its expectation value in $|u_{n\mathbf k}(\mathbf r)\rangle$. This requires that the second term in Eq.~\eqref{Heff} becomes diagonal in the Hilbert space spanned by $|u_{n\mathbf k}(\mathbf r)\rangle$'s. Consequently, we can write
\begin{equation}\label{Eq: L-K condition}
\langle u_{m\mathbf k}(\mathbf r)| \left\{e\mathbf E\cdot\mathrm i\bm\nabla_{\mathbf k}+[\mathcal{H}_0(\mathbf k), \mathcal{S}(\mathbf k)] \right\}|u_{n\mathbf k} (\mathbf r)\rangle= e\mathbf E\cdot \langle u_{m\mathbf k}(\mathbf r)| \mathrm i\bm\nabla_{\mathbf k} |u_{n\mathbf k} (\mathbf r)\rangle+(\varepsilon_{m\mathbf k}-\varepsilon_{n\mathbf k}) \langle u_{m\mathbf k}(\mathbf r)| \mathcal{S}(\mathbf k) |u_{n\mathbf k} (\mathbf r)\rangle=0,    
\end{equation}
where $m\neq n$. Equation~\eqref{Eq: L-K condition} is referred to as the Luttinger-Kohn condition \cite{Luttinger55pr, Schrieffer66pr}, with which the generator of the Schrieffer-Wolff transformation can be solved as
\begin{equation} \label{S}
\mathcal{S}_{mn}(\mathbf k)=\frac{e\mathbf{E}\cdot\mathbfcal{A}_{mn}(\mathbf k)}{\varepsilon_{n\mathbf k}-\varepsilon_{m\mathbf k}}.
\end{equation}
The anti-Hermicity of $\mathcal S(\mathbf k)$ can be checked as $\mathcal S_{nm}^*(\mathbf k)=\tfrac{e\mathbf E\cdot \mathbfcal A_{nm}^*(\mathbf k)}{\varepsilon_{m\mathbf k}-\varepsilon_{n\mathbf k}} = \tfrac{e\mathbf E \cdot \mathbfcal A_{mn}(\mathbf k)}{\varepsilon_{m\mathbf k}-\varepsilon_{n\mathbf k}}=-\mathcal S_{mn}(\mathbf k)$, where  $\mathbfcal A_{nm}^*(\mathbf k)=\mathbfcal A_{nm}(\mathbf k)$ is used. We mention that the diagonal entries of $\mathcal S(\mathbf k)$ cannot be uniquely determined, as the anti-Hermicity only requires the real parts of them to vanish. Such entries are set to zero in Ref.~\cite{YanBH24prl}, but their specific values would not affect the calculation of the electric-field-modified Berry connection and band energy. To see this, we first write down the eigenenergy of $\mathcal{H}(\mathbf k)$, which coincides with the expectation value of $\mathcal{H}_{\text{eff}}(\mathbf k)$, as
\begin{align}
\langle u_{n\mathbf k}&(\mathbf r)|\mathcal{H}_{\text{eff}}(\mathbf k)\left| u_{n\mathbf k}(\mathbf r) \right \rangle = \varepsilon_{n\mathbf k} + \lambda e\mathbf E\cdot\mathbfcal A_n(\mathbf k) + \lambda^2 \left\langle u_{n\mathbf k}(\mathbf r)\left| \left([e\mathbf E\cdot\mathrm i\bm\nabla_{\mathbf k}, \mathcal{S}(\mathbf k)]+\frac{1}{2}[[\mathcal{H}_0(\mathbf k),\mathcal S(\mathbf k)],\mathcal{S}(\mathbf k)]  \right) \right| u_{n\mathbf k}(\mathbf r) \right \rangle \nonumber
\\
&=\varepsilon_{n\mathbf k} + \lambda e\mathbf E\cdot\mathbfcal A_n(\mathbf k) + \lambda^2 \left\langle u_{n\mathbf k}(\mathbf r)\right| [e\mathbf E\cdot\mathrm i\bm\nabla_{\mathbf k}, \mathcal{S}(\mathbf k)]  \left| u_{n\mathbf k}(\mathbf r) \right \rangle  +  \lambda^2 \left\langle u_{n\mathbf k}(\mathbf r)\right| [\varepsilon_{n\mathbf k} \mathcal S^2(\mathbf k) - \mathcal S(\mathbf k) \mathcal H_0(\mathbf k) \mathcal S(\mathbf k)] \left| u_{n\mathbf k}(\mathbf r) \right \rangle \nonumber
\\
&=\varepsilon_{n\mathbf k} + \lambda e\mathbf E\cdot\mathbfcal A_n(\mathbf k) + \lambda^2 \left\langle u_{n\mathbf k}(\mathbf r)\right| [e\mathbf E\cdot\mathrm i\bm\nabla_{\mathbf k}, \mathcal{S}(\mathbf k)]  \left| u_{n\mathbf k}(\mathbf r) \right \rangle  +  \lambda^2 \sum_{m\neq n} (\varepsilon_{n\mathbf k}-\varepsilon_{m\mathbf k}) S_{nm}(\mathbf k) S_{mn}(\mathbf k) \nonumber
\\
&=\varepsilon_{n\mathbf k} + \lambda e\mathbf E\cdot\mathbfcal A_n(\mathbf k) + \lambda^2 \left\langle u_{n\mathbf k}(\mathbf r)\right| [e\mathbf E\cdot\mathrm i\bm\nabla_{\mathbf k}, \mathcal{S}(\mathbf k)]  \left| u_{n\mathbf k}(\mathbf r) \right \rangle  -  \lambda^2 \sum_{m\neq n}  e\mathbf E\cdot \mathbfcal A_{nm}(\mathbf k) S_{mn}(\mathbf k) \nonumber
\\
&=\varepsilon_{n\mathbf k} + \lambda e\mathbf E\cdot\mathbfcal A_n(\mathbf k) + \lambda^2 \sum_{m\neq n} [e\mathbf E\cdot \mathbfcal A_{nm}(\mathbf k) S_{mn}(\mathbf k) - S_{nm}(\mathbf k)  \mathbfcal A_{mn}(\mathbf k) \cdot e\mathbf E] -  \lambda^2 \sum_{m\neq n}  e\mathbf E\cdot \mathbfcal A_{nm}(\mathbf k) S_{mn}(\mathbf k) \nonumber
\\
&=\varepsilon_{n\mathbf k} + \lambda e\mathbf E\cdot\mathbfcal A_n(\mathbf k) + \lambda^2e^2 \sum_{m\neq n} \frac{\mathbf E \cdot  \mathbfcal A_{nm}(\mathbf k)}{\varepsilon_{n\mathbf k}-\varepsilon_{m\mathbf k}} \mathbfcal A_{mn}(\mathbf k) \cdot \mathbf E. 
\end{align}
The electric-field-modified band energy is extracted as the gauge invariant part of $\langle u_{n\mathbf k}(\mathbf r)|\mathcal{H}_{\text{eff}}(\mathbf k)\left| u_{n\mathbf k}(\mathbf r) \right \rangle$ when setting $\lambda=1$. Explicitly, it reads
\begin{equation}\label{Eq: En2-2}
\begin{aligned}
\tilde{\varepsilon}_{n\mathbf k} &=\varepsilon_{n\mathbf k} + e^2 \sum_{m\neq n} \frac{\mathbf E \cdot  \mathbfcal A_{nm}(\mathbf k)}{\varepsilon_{n\mathbf k}-\varepsilon_{m\mathbf k}} \mathbfcal A_{mn}(\mathbf k) \cdot \mathbf E = \varepsilon_{n\mathbf k} + e^2\mathbf E \cdot \Re \sum_{m\neq n} \frac{  \mathbfcal A_{nm}(\mathbf k) \mathbfcal A_{mn}(\mathbf k) }{\varepsilon_{n\mathbf k}-\varepsilon_{m\mathbf k}} \cdot \mathbf E
\\
&= \varepsilon_{n\mathbf k}+\frac{e}{2}\mathbf E \cdot \mathbf G_n(\mathbf k) \cdot \mathbf E.
\end{aligned}    
\end{equation}
Equation~\eqref{Eq: En2-2} is identical to the electric-field-modified band energy derived by the standard perturbation theory [Eq.~\eqref{Eq: en_tot}].

We now turn to evaluate the electric-field-modified Berry connection. With the Schrieffer-Wolff transformed unit-cell-periodic state $\mathrm e^{\mathcal S(\mathbf k)} |u_{n\mathbf k}(\mathbf r)\rangle$, where we set $\lambda=1$ again, the electric-field-modified Berry connection reads
\begin{equation} \label{A_LK}
\begin{aligned}
\tilde{\mathbfcal{A}}_n(\mathbf k)&= \langle u_{n\mathbf k}(\mathbf r)| \mathrm e^{-\mathcal S(\mathbf k)} \mathrm i \bm\nabla_{\mathbf k} \mathrm e^{\mathcal S(\mathbf k)}| u_{n\mathbf k}(\mathbf r) \rangle \simeq \langle u_{n\mathbf k}(\mathbf r)| \{ \mathrm i \bm\nabla_{\mathbf k} + [\mathrm i \bm\nabla_{\mathbf k}, \mathcal S(\mathbf k)] \} | u_{n\mathbf k}(\mathbf r) \rangle 
\\
&=\mathbfcal{A}_n(\mathbf k)+\sum_{m\neq n}[\mathbfcal{A}_{nm}(\mathbf k) \mathcal{S}_{mn}(\mathbf k) - \mathcal{S}_{nm}(\mathbf k)\mathbfcal{A}_{mn}(\mathbf k)]=\mathbfcal{A}_n(\mathbf k)+ 2\Re\sum_{m\neq n}\mathbfcal{A}_{nm}(\mathbf k)\mathcal{S}_{mn}(\mathbf k)
\\
&=\mathbfcal{A}_n(\mathbf k)+2e\text{Re}\sum_{m\neq n}\frac{\mathbfcal{A}_{nm}(\mathbf k)\mathbfcal{A}_{mn}(\mathbf k)}{\varepsilon_{n\mathbf k}-\varepsilon_{m\mathbf k}} \cdot\mathbf E = \mathbfcal{A}_n(\mathbf k)+\mathbf  G_n(\mathbf k)\cdot \mathbf E.
\end{aligned}
\end{equation}
Equation~\eqref{A_LK} is identical to the electric-field-modified Berry connection derived by the standard perturbation theory [Eq.~\eqref{Eq: An_tot}].

\section{Nonlinear Transport}
\label{sec4}
We now evaluate the second-order nonlinear transport. As discussed in the main text, this requires expanding the velocity and non-equilibrium distribution function to the second order of $\mathbf E$. In this section, we will first acquire the expansions and then derive the resulting nonlinear conductivity.

\subsection{Velocity}
\label{sec4a}
The velocity comprises a drift term associated with the electric-field-modified band energy $\tilde \varepsilon_{n\mathbf k}$ and an anomalous term associated with the electric-field-modified Berry curvature $\tilde {\mathbf \Omega}_n(\mathbf k)=\bm \nabla_{\mathbf k} \times \tilde{\mathbfcal A}_n (\mathbf k)$ \cite{xiaodi2010}. Explicitly, the velocity associated with the $n$th energy band reads 
\begin{equation}
\begin{aligned}
\tilde{\mathbf{v}}_n(\mathbf k)&=\frac{1}{\hbar}\bm\nabla_{\mathbf{k}}\tilde{\varepsilon}_{n\mathbf k}-\dot{\mathbf{k}}\times\tilde{\boldsymbol{\Omega}}_n(\mathbf k) =\frac{1}{\hbar}\bm\nabla_{\mathbf{k}}\left[ \varepsilon_{n\mathbf k}+\frac{e}{2} \mathbf E \cdot \mathbf G_n(\mathbf k) \cdot \mathbf E\right]+\frac{e}{\hbar}\mathbf{E}\times \left\{ \bm \nabla_{\mathbf k} \times \left[ \mathbfcal{A}_n(\mathbf k)+\mathbf  G_n(\mathbf k)\cdot \mathbf E \right] \right\}
\\
&=\frac{1}{\hbar}\bm\nabla_{\mathbf{k}}\varepsilon_{n\mathbf k}+\frac{e}{\hbar}\mathbf{E}\times\boldsymbol{\Omega}_n(\mathbf k)+\frac{e}{2\hbar}\bm\nabla_{\mathbf{k}} \left[ \mathbf{E}\cdot\mathbf{G}_n(\mathbf k) \cdot\mathbf{E} \right]+\frac{e}{\hbar}\mathbf{E}\times\left\{ \bm \nabla_{\mathbf{k}}\times \left[\mathbf{G}_n(\mathbf k)\cdot\mathbf{E}\right]\right\},
\end{aligned}
\end{equation}
where we have used the equation of motion $\hbar \dot{\mathbf{k}}=-e\mathbf{E}$. For transparency, it is more convenient to express $\tilde{\mathbf{v}}_n(\mathbf k)$ in the component form as
\begin{equation}\label{Eq: velocity}
\begin{aligned}
\tilde{v}_n^i&=\frac{1}{\hbar}\partial_{i}\varepsilon_n+\frac{e}{\hbar}\epsilon_{ijk}E_j\Omega_n^k+\frac{e}{2\hbar}\partial_{i}G_n^{jk}E_jE_k+\frac{e}{\hbar}\epsilon_{ijk} E_j \epsilon_{klr}\partial_lG_n^{rs}E_s
\\
&=\frac{1}{\hbar}\partial_{i}\varepsilon_n+\frac{e}{\hbar}\epsilon_{ijk}\Omega_n^kE_j+\frac{e}{2\hbar}\partial_{i}G_n^{jk}E_jE_k+\frac{e}{\hbar}(\delta_{il}\delta_{jr}-\delta_{ir}\delta_{jl})\partial_lG_n^{rs}E_jE_s
\\
&=\frac{1}{\hbar}\partial_{i}\varepsilon_n+\frac{e}{\hbar}\epsilon_{ijk}\Omega_n^kE_j+\frac{e}{2\hbar}\partial_{i}G_n^{jk}E_jE_k+\frac{e}{\hbar}\left(\partial_iG_n^{js}-\partial_jG_n^{is}\right)E_jE_s
\\
&=\frac{1}{\hbar}\partial_{i}\varepsilon_n+\frac{e}{\hbar}\epsilon_{ijk}\Omega_n^kE_j+\frac{e}{2\hbar}\left[3\partial_iG_n^{jk}-\left( \partial_jG_n^{ik} + \partial_kG_n^{ij}\right) \right]E_jE_k
\\
&=\frac{1}{\hbar}\partial_{i}\varepsilon_n+\frac{e}{\hbar} \Omega_n^{ij} E_j+\frac{e}{2\hbar}\left[3\partial_iG_n^{jk}-\left( \partial_jG_n^{ik} + \partial_kG_n^{ij}\right) \right]E_jE_k
\\
&=\frac{1}{\hbar}\partial_{i}\varepsilon_n+\frac{e}{2 \hbar} (\Omega_n^{ij} E_j +\Omega_n^{ik} E_k) +\frac{e}{2\hbar}\left[3\partial_iG_n^{jk}-\left( \partial_jG_n^{ik} + \partial_kG_n^{ij}\right) \right]E_jE_k,
\end{aligned}  
\end{equation}
where we have omitted the argument $\mathbf k$ in variables to circumvent misguidance in subscripts/superscripts (e.g., $\varepsilon_{n\mathbf k}\rightarrow \varepsilon_{n}$ and $\partial_{k_i} \rightarrow \partial_i$) and we have also symmetrized indices $\{j,k\}$ (e.g., by defining the Berry curvature tensor $\Omega_n^{ij}=\epsilon_{ijk}\Omega_n^k = \partial_i\mathcal A_n^j-\partial_j\mathcal A_n^i$).

\subsection{Boltzmann Formalism}
\label{sec4b}
The non-equilibrium distribution function $f(\mathbf r,\mathbf k, t)$ can be found using the Boltzmann formalism. In the phase space, the non-equilibrium distribution function is governed by \cite{ashcroft1976}
\begin{equation} \label{boltzmann1}
\begin{aligned}
\frac{\partial f}{\partial t}+\dot{\mathbf{r}}\cdot\bm \nabla_{\mathbf r} f+\dot{\mathbf{k}}\cdot \bm\nabla_{\mathbf k} f=\mathcal{I}\{f\},
\end{aligned}
\end{equation}
where $\mathcal{I}\{f\}$ is the collision integral. For simplicity, we focus on the stationary case such that $f$ does not vary with time, i.e., $f=f(\mathbf r,\mathbf k)$. Moreover, for a weak applied electric field, it is legitimate to assume that $f$ is approximately uniform, resulting in a further simplification $f=f(\mathbf k)$. In the relaxation time approximation, the collision integral can be written as $-[f(\mathbf k)-f_0(\tilde\varepsilon_{n\mathbf k})]/\tau$, where we assume a constant relaxation time $\tau$ for simplicity and $f_0$ is the Fermi-Dirac distribution function. Consequently, Eq.~\eqref{boltzmann1} is reduced to
\begin{equation}
\dot{\mathbf{k}}\cdot \bm\nabla_{\mathbf k} f(\mathbf k)=-\frac{f(\mathbf k)-f_0(\tilde\varepsilon_{n\mathbf k})}{\tau}.
\end{equation}
Making use of the equation of motion $\hbar\dot{\mathbf k}=-e\mathbf E$, the non-equilibrium distribution function admits a power series solution
\begin{equation} \label{f1}
f(\mathbf k)=\left(1-\frac{e\tau}{\hbar}\mathbf{E}\cdot\bm\nabla_\mathbf{k} \right)^{-1} f_0 (\tilde\varepsilon_{n\mathbf k})=\sum_{\nu=0}^\infty \left(\frac{e\tau}{\hbar}\mathbf{E}\cdot\bm\nabla_\mathbf{k} \right)^\nu f_0 (\tilde\varepsilon_{n\mathbf k}).
\end{equation}
Plugging Eq.~\eqref{Eq: en_tot} into Eq.~\eqref{f1}, the non-equilibrium distribution function, to the second order of $\mathbf E$, reads
\begin{equation} \label{f2}
f(\mathbf k)=f_0(\varepsilon_{n\mathbf k}) + \frac{e}{2}\mathbf E \cdot \mathbf G_n(\mathbf k) \cdot \mathbf E \left.\frac{\partial f_0(\varepsilon)}{\partial \varepsilon}\right|_{\varepsilon=\varepsilon_{n\mathbf k}} + \frac{e\tau}{\hbar}\mathbf{E}\cdot\bm\nabla_\mathbf{k}  f_0 (\varepsilon_{n\mathbf k}) + \left( \frac{e\tau}{\hbar}\mathbf{E}\cdot\bm\nabla_\mathbf{k}\right)^2  f_0 (\varepsilon_{n\mathbf k}).
\end{equation}
In a more compact form, we rewrite the non-equilibrium distribution function as
\begin{equation} \label{f3}
f=f_0 + \frac{e}{2}G_n^{ij} E_iE_j f_0' + \frac{e\tau}{\hbar} E_i\partial_i  f_0 +  \frac{e^2\tau^2}{\hbar^2}E_iE_j\partial_i\partial_j  f_0,
\end{equation}
where we use the notations $f= f(\mathbf k)$, $f_0= f_0(\varepsilon_{n\mathbf k})$, and $f_0'=\frac{\partial f_0(\varepsilon)}{\partial \varepsilon}|_{\varepsilon=\varepsilon_{n\mathbf k}}$.

%By considering the correction of energy [Eq. (\ref{Eq: En2-1})], the Fermi distribution function $f_0$ is now a function of electric field $\mathbf{E}$. To obtain the nonlinear current, we expand $f_0$ as a series:
%\begin{equation}\label{Eq：f0_expansion}
%\begin{aligned}
%f_0(\varepsilon_n+\frac{e}{2}\mathbf{E}\cdot \mathbf{G}_n\cdot\mathbf{E})\simeq f_0(\varepsilon_n)+\frac{e}{2}\mathbf{E}\cdot \mathbf{G}_n\cdot\mathbf{E}f_0'.
%\end{aligned}
%\end{equation}
%This is a subtle clue to examine the distinct expressions in the Ref. \cite{kaplan2024} and Refs. \cite{gaoyang2014,wangchong2021,liuhuiying2021}. 

\subsection{Nonlinear Conductivity}
\label{sec4c}
With the velocity and the non-equilibrium distribution function in hand, it is straightforward to calculate the nonlinear conductivity. For this purpose, we first evaluate the current density
\begin{equation}\label{Eq: e-Current}
\begin{aligned}
\mathbf{J}=&\frac{-e}{\mathcal{V}}\sum_n \sum_{\mathbf{k}} \tilde {\mathbf v}_n(\mathbf k) f(\mathbf k)=-e \sum_n \int [d\mathbf{k}] \tilde {\mathbf v}_n(\mathbf k) f(\mathbf k),\\
\end{aligned} 
\end{equation}
where $\mathcal V$ is the volume of the system and $[d\mathbf{k}]\equiv d^d\mathbf{k}/(2\pi)^d$ with dimension $d$. As we have respectively found $\tilde {\mathbf v}_n(\mathbf k)$ and $f(\mathbf k)$ to the second order of $\mathbf E$ in Secs.~\ref{sec4a} and~\ref{sec4b}, Eq.~\eqref{Eq: e-Current} can also be evaluated to the second order of $\mathbf E$. In the component form, the current density can thus be written as
\begin{equation}
J_i\simeq\sigma_{ij}E_j+\sigma_{ijk}E_j E_k,
\end{equation}
where $\sigma_{ij}$ is the linear conductivity and $\sigma_{ijk}$ is the second-order nonlinear conductivity. In the rest of this section, we will focus on the latter. It is worth noting that the nonlinear transport can be divided into three categories according to the dependence on the relaxation time $\tau$. First, the $\tau^2$-dependent nonlinear current reads
\begin{equation} \label{Jd}
J_{i}^{\text{d}}=-e\sum_n\int [d\mathbf{k}] v_n^i \frac{e^2\tau^2}{\hbar^2}E_jE_k\partial_j\partial_k  f_0 =-\frac{\tau^2 e^3}{\hbar^3} \sum_n\int [d\mathbf{k}] (\partial_{i}\varepsilon_n)  E_j E_k \partial_j \partial_k f_0 =-\frac{e^3\tau^2}{\hbar^3}\sum_n\int [d\mathbf{k}] \big(\partial_i\partial_j  \partial_k\varepsilon_n\big)f_0E_jE_k,
\end{equation}
which is associated with the Drude mechanism. Second, the $\tau$-dependent nonlinear current reads
\begin{equation} \label{Jbc}
\begin{split}
J_{i}^{\text{bc}}&=-e\sum_n\int [d\mathbf{k}] \frac{e}{2\hbar}(\Omega_n^{ij}E_j+\Omega_n^{ik}E_k)  \frac{e\tau}{\hbar} E_l\partial_l f_0 =- \frac{e^3\tau}{2\hbar^2} \sum_n\int [d\mathbf{k}] (\Omega_n^{ij}E_j E_k\partial_k f_0 +\Omega_n^{ik}E_k E_j\partial_j f_0 )   
\\
& =\frac{e^3\tau}{2\hbar^2} \sum_n\int [d\mathbf{k}] (\partial_k \Omega_n^{ij} + \partial_j \Omega_n^{ik} )f_0 E_j E_k,    
\end{split}
\end{equation}
which is referred to as the Berry curvature dipole contribution. Lastly, the $\tau$-free (i.e., intrinsic) nonlinear current reads
\begin{equation}\label{Jqm}
\begin{aligned}
J_{i}^{\text{qm}}&=-e\sum_n\int [d\mathbf{k}] v_n^i \frac{e}{2}G_n^{jk} E_jE_k f_0' -e\sum_n\int [d\mathbf{k}] \frac{e}{2\hbar}\left[3\partial_iG_n^{jk}-\left( \partial_jG_n^{ik} + \partial_kG_n^{ij}\right) \right]E_jE_k f_0
\\
&=-\frac{e^2}{2\hbar}\sum_n\int [d\mathbf{k}] (\partial_i\varepsilon_n) G_n^{jk} E_jE_k f_0' -\frac{e^2}{2\hbar}\sum_n\int [d\mathbf{k}] \left[3\partial_iG_n^{jk}-\left( \partial_jG_n^{ik} + \partial_kG_n^{ij}\right) \right]E_jE_k f_0
\\
&=\frac{e^2}{2\hbar}\sum_n\int [d\mathbf{k}]  (\partial_iG_n^{jk}) E_jE_k  f_0 -\frac{e^2}{2\hbar}\sum_n\int [d\mathbf{k}] \left[3\partial_iG_n^{jk}-\left( \partial_jG_n^{ik} + \partial_kG_n^{ij}\right) \right]E_jE_k f_0
\\
&= -\frac{e^2}{\hbar}\sum_n\int [d\mathbf{k}] \left[\partial_iG_n^{jk}-\frac{1}{2} \left( \partial_jG_n^{ik} + \partial_kG_n^{ij}\right) \right] f_0 E_jE_k.
\end{aligned}
\end{equation}
which is known as the quantum metric dipole contribution and will be further demonstrated in the following [see Eq.~\eqref{qmd}].

The nonlinear conductivities associated with the Drude, Berry curvature dipole, and quantum metric dipole mechanisms can be respectively read off from Eqs.~\eqref{Jd},~\eqref{Jbc}, and~\eqref{Jqm} as
\begin{align}
&\sigma_{ijk}^{\text{d}}=-\frac{e^3\tau^2}{\hbar^3}\sum_n\int [d\mathbf{k}] \big(\partial_i\partial_j  \partial_k\varepsilon_n\big)f_0, \label{d_f}
\\
&\sigma_{ijk}^{\text{bc}}=\frac{e^3\tau}{2\hbar^2}\sum_n\int [d\mathbf{k}]\big(\partial_k\Omega_n^{ij}+\partial_j\Omega_n^{ik}\big)f_0, \label{bc_f}
\\
&\sigma_{ijk}^{\text{qm}}=-\frac{e^2}{\hbar}\sum_n\int [d\mathbf{k}]\left[\partial_i G_n^{jk}-\frac{1}{2}\left(\partial_k  G_n^{ij}+\partial_j G_n^{ik}\right)\right]f_0, \label{qm_f}
\end{align}
which correspond to Eqs.~(\textcolor{red}{11})-(\textcolor{red}{13}) in the main text. Alternatively, through integrating by parts, these expressions can be rewritten as
\begin{align}
&\sigma_{ijk}^{\text{d}}=\frac{e^3\tau^2}{\hbar^2}\sum_n\int [d\mathbf{k}] v_n^i \big( \partial_j  \partial_k\varepsilon_n\big)f_0', \label{d_fp}
\\
&\sigma_{ijk}^{\text{bc}}=-\frac{e^3\tau}{2\hbar}\sum_n\int [d\mathbf{k}]\left(v_n^{k}\Omega_n^{ij}+v_n^j\Omega_n^{ik}\right)f_0', \label{bc_fp}
\\
&\sigma_{ijk}^{\text{qm}}=e^2\sum_n\int [d\mathbf{k}]\left[v_n^i  G_n^{jk}-\frac{1}{2}\left(v_n^k G_n^{ij}+v_n^j G_n^{ik}\right)\right]f_0'. \label{qm_fp}
\end{align}
The derived quantum-metric-induced nonlinear conductivity [Eqs.~\eqref{qm_f} and~\eqref{qm_fp}] is consistent with those listed in Refs.~\cite{gaoyang2014, wangchong2021, liuhuiying2021}, which are not symmetrized for indices $\{j, k\}$. To better illustrate the quantum metric origin of $\sigma_{ijk}^{\text{qm}}$, we rewrite Eqs.~\eqref{qm_f} and~\eqref{qm_fp} as
\begin{align}
\sigma_{ijk}^{\text{qm}} &=-\frac{e^3}{\hbar}\sum_n\int [d\mathbf{k}]\left[\partial_i \mathcal G_n^{jk}-\frac{1}{2}\left(\partial_k  \mathcal G_n^{ij}+\partial_j \mathcal G_n^{ik}\right)\right]f_0,
\\
\sigma_{ijk}^{\text{qm}} &=e^3\sum_n\int [d\mathbf{k}]\left[v_n^i  \mathcal G_n^{jk}-\frac{1}{2}\left(v_n^k \mathcal G_n^{ij}+v_n^j \mathcal G_n^{ik}\right) \right]f_0', \label{kernel}
\end{align}
where $\mathbfcal G_n(\mathbf k)=2\Re\sum_{m\neq n}{\color{black}\mathbfcal A_{nm}(\mathbf k)} \mathbfcal A_{mn}(\mathbf k)/(\varepsilon_{n\mathbf k}-\varepsilon_{m\mathbf k})$ is the band-normalized quantum metric, differing from the Berry connection polarizability $\mathbf G_n(\mathbf k)$ by a factor of $e$. The kernel of Eq.~\eqref{kernel} reads
\begin{equation} \label{qmd}
\Lambda_n^{ijk}(\mathbf k) = v_n^i  \mathcal G_n^{jk}-\frac{1}{2}\left(v_n^k \mathcal G_n^{ij}+v_n^j \mathcal G_n^{ik}\right), 
\end{equation}
which is referred to as the ``quantum metric dipole'' as the band-normalized quantum metric $\mathbfcal G_n(\mathbf k)$ is related to the quantum metric $\mathbf g_n (\mathbf k)=\Re\sum_{m\neq n} \mathbfcal A_{nm}(\mathbf k) \mathbfcal A_{mn}(\mathbf k)$ through $\mathbfcal G_n(\mathbf k)=-\partial \mathbf g_n (\mathbf k)/\partial \varepsilon_{n\mathbf k}$ \cite{note2}.

%
%as the Berry connection polarizability $\mathbf G_n(\mathbf k)$ only differs from the band-normalized quantum metric  by a factor of $e$.}

\bibliographystyle{apsrev}
\bibliography{supp.bib}